\documentclass[epj]{svjour}

\usepackage[OT1]{fontenc}
\usepackage[usenames,dvipsnames]{color}
\usepackage[latin1]{inputenc}
\usepackage[english]{babel}
\usepackage{graphicx}
\usepackage{color}

\usepackage[Gray,squaren]{SIunits}
\usepackage{xspace}
\usepackage{upgreek}
\usepackage{ulem}
\usepackage{epstopdf}
\usepackage{amssymb,amsmath,verbatim,ulem}
\usepackage{enumerate}

\newcommand{\ket}[1]{|{#1}\rangle}
\newcommand{\bra}[1]{\langle{#1}|}

\begin{document}

\title{Nature of interference between Autler-Townes peaks in generic multi-level system}
\author{Elijah Ogaro Nyakang'o\inst{1}, Dangka Shylla\inst{1}, Kirthanaa Indumathi\inst{1} \and Kanhaiya Pandey\inst{1}
\thanks{\emph{email address: kanhaiyapandey@iitg.ac.in} }
}
\institute{Department of Physics, Indian Institute of Technology Guwahati, Guwahati, Assam 781039, India}

\date{\today{}}

\abstract{In this work we present a theoretical frame work to identify the role and the nature of interference between Autler-Townes (AT) peaks (or dressed states) in generic multi-level system. In three-level system, the two AT peaks interferes pair-wise with each other, almost similar to the two-slit interference. In four-level system, the interference between the three AT peaks is also pair-wise only with no higher order of interference analogous to three-slit interference but has a bit more complicated nature of interference. However, in many practical situations in atomic systems only the simple form of interference similar to three-level system dominates. In the three-level system, the nature of interference (i.e. constructive, destructive or no interference) between the two AT peaks  is purely determined by the natural decay rate of the states coupled by the control laser(s). However, in four-level system the nature of interference between the two extreme AT peaks can be tuned from constructive to destructive by tuning the power of the control lasers.}
    
\maketitle

\section{Introduction}

Modification of a weak probe laser absorption by a strong control laser in three-level system is due to two closely related phenomena, one is electromagnetically induced transparency (EIT) and the other is Autler-Townes (AT) splitting. AT absorption peaks are due to the dressed states created by the control laser(s). In this paper we will use both terminologies i.e., AT peaks and dressed states which are basically the same.  EIT in three-level system is AT splitting plus the interference \cite{IMA89} between them and has been addressed theoretically \cite{LIX95,FIM05,SHF96,AGA97,ABT10,TAH14,KBN16} and experimentally using Akaike information criteria in real atomic system \cite{ADS11,GVS13,HJX18,LWZ20} as well as in artificial atomic system \cite{POC14}. The distinction between EIT and AT splitting has also been investigated based upon the quantum memory \cite{ETA18,RSH19}.

In three-level system, the strong control laser creates two AT peaks (dressed states) which are probed by a weak probe laser. The interference between the two AT peaks can be constructive, destructive or no interference depending upon the decay rate of the bare atomic states coupled by the control laser. Further in three-level systems, there are only two dressed states, and hence there is possibility of pair-wise interference only similar to the two-slit interference. However, it is very interesting to investigate the nature of interference with more than two dressed states in a similar fashion to the three-slit interference \cite{SCJ10,ASQ15,SAS15,LZC20}. To probe the nature of interference between three or more dressed states, we consider four-level system or other higher levels. The four-level system or other higher levels, have been extensively studied for various applications \cite{JOX03,BMW08,CXH09,SYK11,HCN05,JSA12,BJX11,GIM18,SLB13,LPR18,HMW14,PKP16,LHF05}, but the role and nature of interference has not been addressed.

The absorption of probe laser in the presence of the control laser(s) in multilevel systems is generally analyzed using two approaches. One is laser induced coherence between the levels (i.e. density matrix approach in bare states), also known as transfer of coherence (TOC) since simultaneous driving of different levels with lasers induces coherence between levels which are not directly driven. The TOC approach is further analyzed by qualitative dressed state approach to explain the probe laser absorption profile. The other approach is dressed states created by the control laser(s) and their excitation by the probe laser. The complete dressed state analysis to derive the exact probe absorption profile is the key to identify the exact nature of interference between AT peaks including some non-trivial aspects. In order to study the nature of interference between AT peaks, we first provide a generic theoretical frame work for probe absorption in the dressed state picture and verify it using TOC approach i.e. density matrix formalism in bare atomic state picture. In the derived formula of probe absorption using dressed state approach, the nature of interference between AT peaks is identified and absorption of the probe and AT peaks plotted for a variety of systems.

\section{The theoretical model}

\subsection{Dressed state approach}

\begin{figure}
   \begin{center}
      \includegraphics[width =0.8\linewidth]{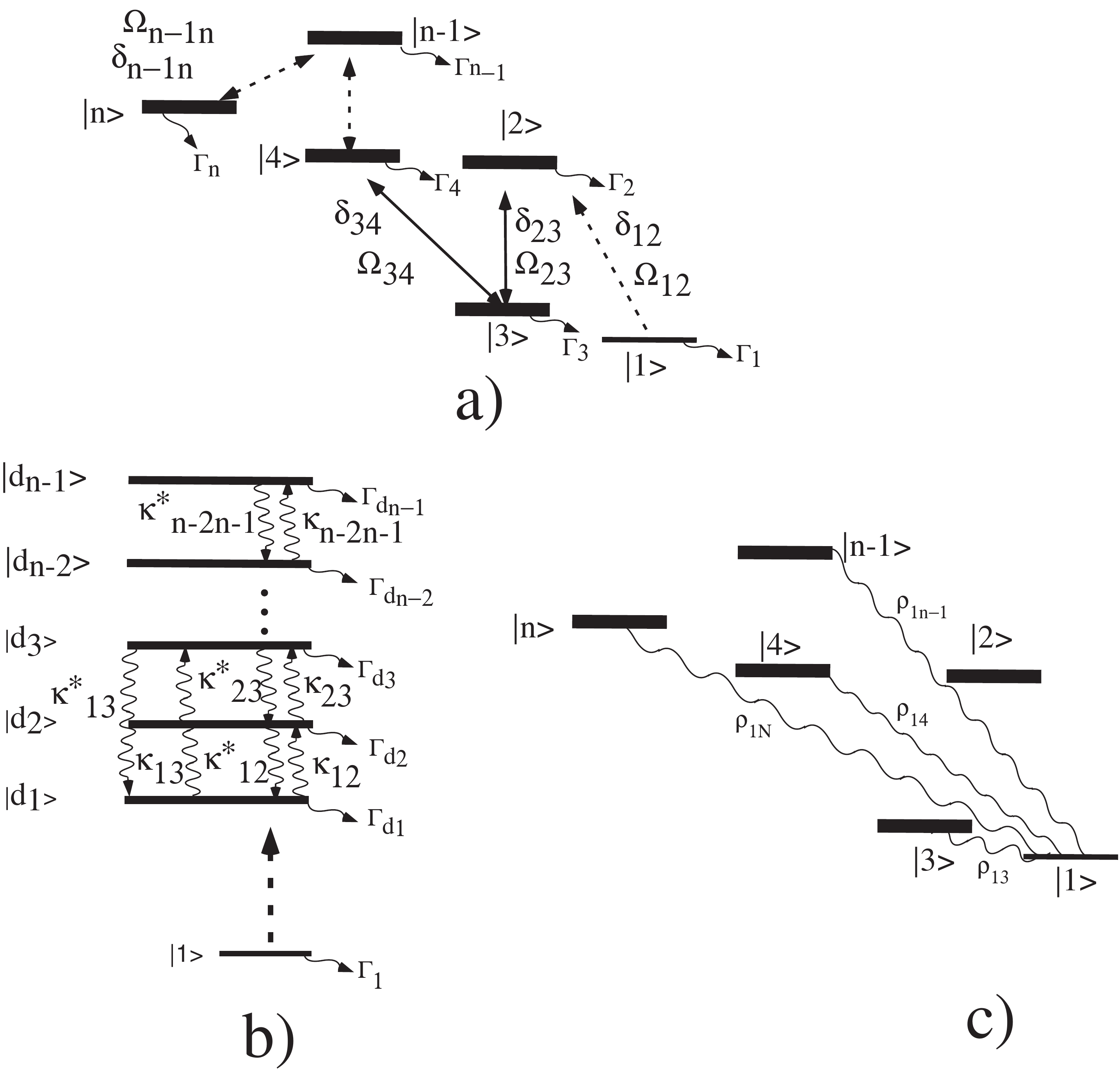}
      \caption {The general n-level atomic system: a)  Bare atomic state picture. b)  Dressed state picture. c) Transfer of coherence shown by curly line between various levels not directly driven by the lasers}
      \label{Fig1}
   \end{center}
\end{figure}

We consider the generic system as shown in Fig. \ref{Fig1}, in which a weak probe laser is driving the transition $\ket{1}$ $\rightarrow$ $\ket{2}$ with Rabi frequency $\Omega_{12}$ and the detuning $\delta_{12}$ (note, the detuning is defined as $\delta_{i-1,i}=\omega^{l}_{i-1,i}\mp(\omega_{i}-\omega_{i-1})$ if $\omega_{i}>\omega_{i-1}$ and $\omega_{i}<\omega_{i-1}$ respectively, where $\omega^{l}_{i-1,i}$ is the frequency of the laser, $\omega_{i}$ and $\omega_{i-1}$ are the frequency of the energy levels). The strong control lasers are driving the transitions $\ket{2}$ $\leftrightarrow$ $\ket{3}$, $\ket{3}$ $\leftrightarrow$ $\ket{4},\cdots,\ket{n-1}$ $\leftrightarrow$ $\ket{n}$ with detunings $\delta_{23}$, $\delta_{34},\cdots,\delta_{n-1 n}$ and the Rabi frequencies $\Omega_{23}$, $\Omega_{34},\cdots,\Omega_{n-1 n}$. The Hamiltonian associated only with the control lasers in the rotating frame with rotating wave approximation is given as
\begin{align}
\label{eq1}
H_c&=-\hbar\delta_{23}\ket{3}\bra{3}-\hbar(\delta_{23}\pm\delta_{34})\ket{4}\bra{4}-\cdots \nonumber\\ 
&-\hbar(\delta_{23}\pm\delta_{34}\pm\cdots\pm\delta_{n-1,n})\ket{n}\bra{n}+\Big\{\frac{\hbar\Omega_{23}}{2}\ket{2}\bra{3} \nonumber\\
&+\frac{\hbar\Omega_{34}}{2}\ket{3}\bra{4}+\cdots+\frac{\hbar\Omega_{n-1n}}{2}\ket{n-1}\bra{n}+h.c.\Big\}
\end{align}  
The sign $\pm$ of $\delta_{i-1,i}$ is chosen according to the level structure. If the energy of the state $\ket{i}$ is higher than the state $\ket{i-1}$, then $\delta_{i-1,i}$ will have + sign and if the energy of the state $\ket{i}$ is lower than the state $\ket{i-1}$, then $\delta_{i-1,i}$ will have - sign. For example, if the two control lasers form a ladder system, the two photon detuning is given as $\delta_{23}+\delta_{34}$ and if they form a lambda system the two photon detuning is $\delta_{23}-\delta_{34}$. Similarly the two photon detuning of a Vee system (which is an inverted lambda system) is also $\delta_{23}-\delta_{34}$. This in general gives $\delta_{23}\pm\delta_{34}$, hence the $\pm$ sign. The eigenvalues ($E_{d_i}=\hbar\Delta_{d_i}$) of the $H_c$ determines the position of the dressed states ($\ket{d_i}$) or AT absorption peaks. The eigenvectors will determine the dressed states which is a linear combination of the bare atomic states.

The transformation from the bare states, $\ket{\Psi_{B}}$ (i.e. $\ket{1},\ket{2},\cdots,\ket{n}$) to dressed state (similar to Morris-Shore transformation \cite{MOS83}), $\ket{\Psi_{D}}$ is given by the unitary matrix $U$ of dimension ${n-1}\times{n-1}$ i.e.,
\begin{equation}
\label{eq2}
\ket{\Psi_{D}}=\hat{U}\ket{\Psi_{B}}
\end{equation}
The general element $ij$ ($i=1,\cdots,{n-1}$, $j=2,\cdots,n$) of this unitary matrix is $u_{ij}$ and the generic form of the $i^{th}$ ($i=1,\cdots,{n-1}$) dressed state $\ket{d_{i}}$ is
\begin{align}
\label{eq3}
\ket{d_i}=u_{i2}\ket{2}+u_{i3}\ket{3}+\cdots+u_{in}\ket{n}
\end{align}
If the total decay rate of the bare atomic state $\ket{i}$ is $\Gamma_{i}$, the decay matrix $\hat{\Gamma}$ in the bare state can easily be written as,
\begin{equation}
\label{eq4}
\hat{\Gamma}=\sum_{i=2}^{n}\frac{\Gamma_{i}}{2}\ket{i}\bra{i}
\end{equation}
This decay matrix, $\hat{\Gamma}$ will transform in the dressed state basis as $\hat{U}\hat{\Gamma}\hat{U}^{\dagger}$. The general ${ij}^{th}$ element of this matrix, $\kappa_{ij}(= \bra{i}\hat{U}\hat{\Gamma}\hat{U}^{\dagger}\ket{j})$ will be,
\begin{align}
\label{eq5}
\kappa_{ij}=\frac{1}{2}(u_{i2}^{\ast}u_{j2}\Gamma_2+u_{i3}^{\ast}u_{j3}\Gamma_3+\cdots+u_{in}^{\ast}u_{jn}\Gamma_{n})
\end{align}
The diagonal elements, $\kappa_{ii}(=\Gamma_{d_{i}})$ of this matrix corresponds to incoherent decay of the dressed states $\ket{d_i}$ and contributes to its linewidth.
\begin{align}
\label{eq6}
\Gamma_{d_i}=\frac{1}{2}(|u_{i2}|^2\Gamma_2+|u_{i3}|^2\Gamma_3+\cdots+|u_{in}|^2\Gamma_{n})
\end{align}
\textit{The coherent decay terms $\kappa_{ij}$ gives rise to interference between the dressed states}. Note that if $\Gamma_{i}(=\Gamma)$ is the same for all bare states, $\kappa_{ij}= \bra{i}\hat{U}\hat{\Gamma}\hat{U}^{\dagger}\ket{j}=\frac{\Gamma}{2}\bra{i}\hat{U}\hat{1}\hat{U}^{\dagger}\ket{j}=0$ $\textrm{for}~i\ne j$. In this case there will be no interference between the dressed states. This might not be clearly evident in the bare state density matrix approach.

The dressed states $\ket{d_i}$ couples with $\ket{1}$ through the probe laser with a coupling strength $\Omega_{p_i}=-\bra{1}\vec{D}.\vec{E}^0_{12}\ket{d_i}/\hbar$ where, $\vec{E}^0_{12}$ is the electric field amplitude associated with the probe laser and $\vec{D}$ is the electric dipole moment operator. With the rotating wave approximation $\bra{1}\vec{D}.\vec{E}^0_{12}\ket{d_i}/\hbar=\bra{1}\vec{D}.\vec{E}^0_{12}u_{i2}\ket{2}/\hbar$ and hence $\Omega_{p_i}=u_{i2}\Omega_{12}$. The amplitude of the excitation path for AT peaks corresponding to the dressed states $\ket{d_i}$ will be proportional to the $\Omega_{p_i}$. The amplitude for the probe absorption corresponding to this peak will be proportional to $|\Omega_{p_i}|^2$.

The equation of motion for the dressed states $\ket{d_i}$ and the bare state $\ket{1}$ is
\scriptsize{}
\begin{equation}
\label{eq7}
i\frac{d}{dt}
\begin{bmatrix}
&C_1 \\
 &C_{d_1}\\
&.\\
&.\\
&C_{d_{n-1}}
\end{bmatrix}
=
\begin{bmatrix}
0 &\frac{\Omega_{p_1}}{2} & \frac{\Omega_{p_2}}{2}&..& \frac{\Omega_{p_{n-1}}}{2} \\
\frac{\Omega^*_{p_1}}{2} &-i\gamma_{d_1}&-i\kappa_{12}& ..&-i\kappa_{1n-1} \\
&.\\
&.\\
\frac{\Omega^*_{p_{n-1}}}{2}&-i\kappa^*_{1n}&-i\kappa^*_{2n-1}&..&-i\gamma_{d_{n-1}}\\
\end{bmatrix}
\begin{bmatrix}
&C_1 \\
 &C_{d_1}\\
&.\\
&.\\
&C_{d_{n-1}}
\end{bmatrix}
\end{equation}
\normalsize{}where, $\gamma_{d_i}=\frac{\Gamma_1}{2}+\Gamma_{d_i}+i\delta_{d_i}$ with $\delta_{d_i}=\delta_{12}+\Delta_{d_i}$. We consider the steady state for the dynamics of all the dressed states $\ket{d_i}$ i.e. $\frac{dC_{d_{i}}}{dt}=0$. The absorption of the probe laser is $\frac{d|C_1|^2}{dt}=\frac{dC_1^*}{dt}C_1+\frac{dC_1}{dt}C_1^*$. For the weak probe we consider $C_1\approx1$ and probe absorption is now given as $\frac{d|C_1|^2}{dt}=\frac{dC_1^*}{dt}+\frac{dC_1}{dt}$. The normalized absorption of the probe laser is given as $-(\Gamma_1+\Gamma_2)/(2|\Omega_{12}|^2)(\frac{dC_1^*}{dt}+\frac{dC_1}{dt})$ so that in the absence of the control lasers, probe absorption is $1$ at resonance ($\delta_{12}=0$).

\subsection{Bare state TOC approach}

\begin{figure}
   \begin{center}
      \includegraphics[width =0.8\linewidth]{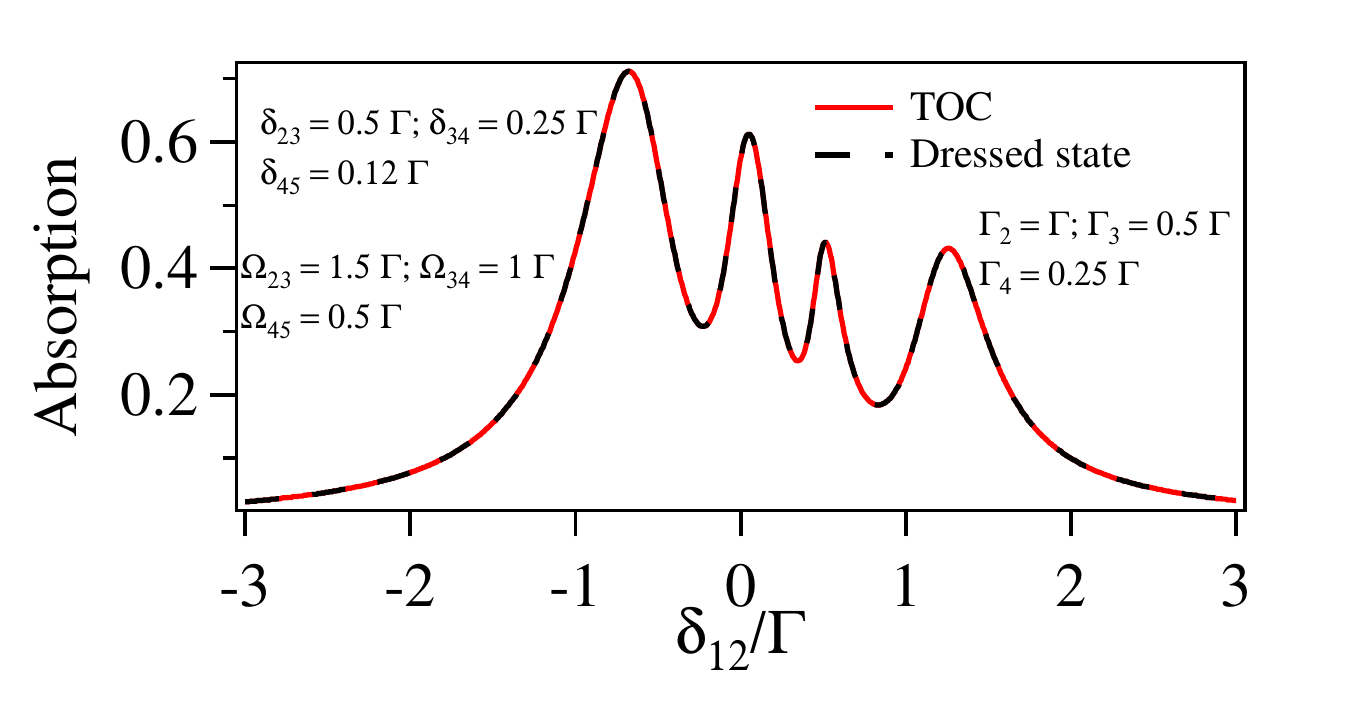}
      \caption{Comparison of the probe absorption obtained using TOC and dressed state approach.}
      \label{Fig2}
   \end{center}
\end{figure}

The other approach which is commonly used to analyze the absorption of the probe laser in a multilevel system is the transfer of coherence (TOC) between the bare states using density matrix formalism and is very well described in our previous work \cite{PAN13}. The absorption of the probe laser is written in terms of the density matrix element between level $\ket{1}$ and $\ket{2}$ i.e. $\rho_{12}$. The solution for $\rho_{12}$ is written in the continued fractional form given below,
\begin{align}
\label{eq8}
\rho_{12}=&\cfrac{\frac{i}{2}\frac{\Omega_{12}}{\gamma_{12}}}
{1+\cfrac{\frac{1}{4}\frac{\mid\Omega_{23}\mid^2}{\gamma_{12}\gamma_{13}}}
{1+\cfrac{\frac{1}{4}\frac{\mid\Omega_{34}\mid^2}{\gamma_{13}\gamma_{14}}}
{1+\cfrac{\frac{1}{4}\frac{\mid\Omega_{45}\mid^2}{\gamma_{14}\gamma_{15}}}
{1+\cfrac{\frac{1}{4}\frac{\mid\Omega_{56}\mid^2}{\gamma_{15}\gamma_{16}}}
{1+\cfrac{.}{1+\cfrac{.}{1+\frac{1}{4}\frac{\mid\Omega_{n-1 n}\mid^2}{\gamma_{1 n-1}\gamma_{1 n}}}}}}}}}
\end{align}
where,
\begin{equation}
\label{eq9}
\gamma_{1j}=\frac{\Gamma_{1}+\Gamma_{j}}{2}-i\displaystyle\sum_{i=1}^{j-1}(-1)^{i+1}\delta_{i,i+1}
\end{equation}
is the decoherence rate of transferred coherence between $\ket{1}$ and $\ket{j}$ of $\rho_{1j}$. The normalized absorption of the probe laser is given as $\frac{\Gamma_1+\Gamma_2}{\Omega_{12}}\rho_{12}$ so that absorption of the resonant probe laser is 1 in the absence of the control lasers. The validity of the dressed state approach is tested by comparing the probe absorption in the two approaches. The absorption of the probe completely match in the two approaches for various number of levels and of various type of systems for different atomic and control laser parameters. For example, the probe absorption of the five-level system for the parameters given in the annotation is as shown in Fig. \ref {Fig2}.

\section{The nature of interference in the multi-level systems}

\subsection{Three-level system}

\begin{figure}
   \begin{center}
      \includegraphics[width =0.8\linewidth]{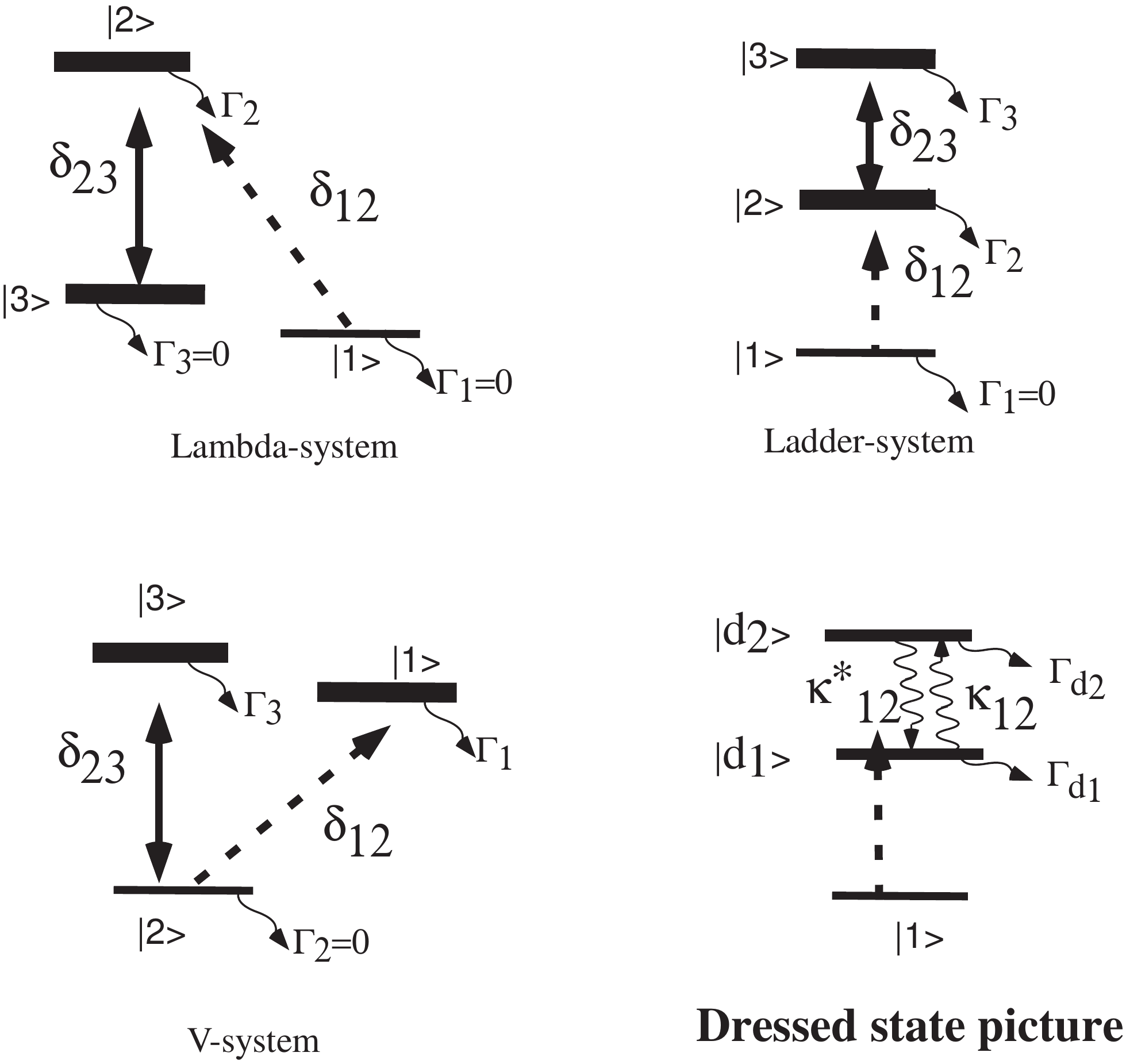}
      \caption{Three level atomic system: a)  Bare atomic state picture. b)  Dressed state picture.}
      \label{Fig3}
   \end{center}
\end{figure}

The nature of interference in the three-level system has been addressed in earlier works \cite{LIX95,FIM05,SHF96,AGA97,ABT10,TAH14,KBN16}, however, we present this system here for completeness and also in a comprehensive way for the rest of the work. We consider a three-level system as shown in Fig. \ref{Fig3} in which a strong control laser is driving the transition $\ket{2}\leftrightarrow\ket{3}$ and a weak probe laser is driving the transition $\ket{1}\rightarrow\ket{2}$. The detuning of the probe laser and the control laser is $\delta_{12}$ and $\delta_{23}$ respectively in the bare atomic state picture. The Rabi frequency of the probe laser is $\Omega_{12}$ and the control laser is $\Omega_{23}$. The strong control laser creates dressed states whose position is determined by the eigenvalues of the Hamiltonian associated with the control laser,
\begin{equation}
\label{eq10}
H_c=\hbar
\begin{bmatrix}
0 &\frac{\Omega_{23}}{2} \\
\frac{\Omega^*_{23}}{2} &-\delta_{23}
\end{bmatrix}
\end{equation}

Using Eq. \ref{eq7}, the equation of motion for the coefficients of $\ket{1}$, $\ket{d_1}$ and $\ket{d_2}$ are given as, \small{}
\begin{align}
\label{eq11}
i\frac{dC_1}{dt}&=\frac{\Omega_{p_1}}{2}C_{d_1}+\frac{\Omega_{p_2}}{2}C_{d_2}\nonumber\\
i\frac{dC_{d_1}}{dt}&=\frac{\Omega^*_{p_1}}{2}C_1-i\gamma_{d_1}C_{d_1}-i\kappa_{12}C_{d_2}\\
i\frac{dC_{d_2}}{dt}&=\frac{\Omega^*_{p_2}}{2}C_1-i\kappa^*_{12}C_{d_1}-i\gamma_{d_2}C_{d_2}\nonumber
\end{align}
\normalsize{}Considering the steady state case for the time evolution of the dressed states i.e. $\frac{dC_{d_1}}{dt}=\frac{dC_{d_2}}{dt}=0$, we get the following equation which is proportional to probe absorption.
\small{}
\begin{equation}
\label{eq12}
\frac{dC_1}{dt}=-\underbrace{\frac{1/4}{1-\frac{|\kappa_{12}|^2}{\gamma_{d_1}\gamma_{d_2}}}}_{\textrm{Normalization}}\left[\underbrace{\frac{|\Omega_{p_1}|^2}{\gamma_{d_1}}+\frac{|\Omega_{p_2}|^2}{\gamma_{d_2}}}_{\textrm{AT peaks}}-\underbrace{\frac{\kappa_{12}\Omega^*_{p_1}\Omega_{p_2}+c.c}{\gamma_{d_1}\gamma_{d_2}}}_{\textrm{Interference}}\right]
\end{equation}
\normalsize{}The eigenvalues of the two dressed states $\ket{d_1}$ and $\ket{d_2}$ corresponding to the Hamiltonian in Eq. \ref{eq10} and the various parameters are: 
\small{}
\begin{align}
\label{eq13}
&E_{d_1}=\frac{\hbar}{2}[-\delta_{23}-\sqrt{\delta^2_{23}+\Omega^2_{23}}]; \ket{d_1}=-\cos{\theta}\ket{2}+\sin{\theta}\ket{3}\nonumber\\
&\Gamma_{d_1}=\frac{1}{2}(\cos^2{\theta}\Gamma_2+\sin^2{\theta}\Gamma_3);\nonumber\\
&E_{d_2}=\frac{\hbar}{2}[-\delta_{23}+\sqrt{\delta^2_{23}+\Omega^2_{23}}];\ket{d_2}=\sin{\theta}\ket{2}+\cos{\theta}\ket{3}\nonumber\\
&\Gamma_{d_2}=\frac{1}{2}(\sin^2{\theta}\Gamma_2+\cos^2{\theta}\Gamma_3); \nonumber\\
&\kappa_{12}=-\frac{1}{4}(\Gamma_2-\Gamma_3)\sin{2\theta}; \Omega_{p_1}=-\Omega_{12}\cos\theta;\nonumber\\
& \Omega_{p_2}=\Omega_{12}\sin{\theta};\\
&\textrm{where,} \tan{2\theta}=-\frac{\Omega_{23}}{\delta_{23}}\nonumber
\end{align}

\normalsize{}The nature of interference between the two dressed states $\ket{d_1}$ and $\ket{d_2}$ is determined by the parameter $\kappa_{12}$. It is possible that the interference parameter, $\kappa_{12}$ can change its sign for given $\Gamma_2$ and $\Gamma_3$ by tuning the control laser detuning, $\delta_{23}$ from positive to negative and vice versa. However, the amplitude of excitation, $\Omega_{p_2}$ of the dressed state $\ket{d_2}$ also changes its sign as the control laser detuning $\delta_{23}$ is changed. The overall interference term $\kappa_{12}\Omega_{p_1}\Omega_{p_2}$ will not change sign and it is therefore not possible to change the sign of interference by changing the detuning of the control laser.

Eq. \ref{eq12} can also be written in terms of $\theta$ parameters given in Eq. \ref{eq13} as follows,  
\begin{equation}
\label{eq14}
\frac{dC_1}{dt}=-\underbrace{\frac{|\Omega_{12}|^2/4}{1-\frac{|\kappa_{12}|^2}{\gamma_{d_1}\gamma_{d_2}}}}_{\textrm{Normalization}}\left[\underbrace{\frac{\cos^2{\theta}}{\gamma_{d_1}}+\frac{\sin^2{\theta}}{\gamma_{d_2}}}_{\textrm{AT peaks}}-\underbrace{\frac{\Gamma_2-\Gamma_3}{4}\frac{\sin^2{2\theta}}{\gamma_{d_1}\gamma_{d_2}}}_{\textrm{Interference}}\right]
\end{equation}
For the three-level system, Eq. \ref{eq8} for TOC approach is reduced as follows,
\begin{eqnarray}
\label{eq15}
\begin{split}
&\rho_{12}= \cfrac{\frac{i}{2}\frac{\Omega_{12}}{\gamma_{12}}}
{1+\frac{1}{4}\frac{\mid\Omega_{23}\mid^2}{\gamma_{12}\gamma_{13}}}
\end{split}
\end{eqnarray}
The comparison of probe absorption using Eq. \ref{eq14} (or Eq. \ref{eq12}) and Eq. \ref{eq15} for the two approaches shows a complete match for the various parameters.
\begin{figure}
   \begin{center}
      \includegraphics[width =0.8\linewidth]{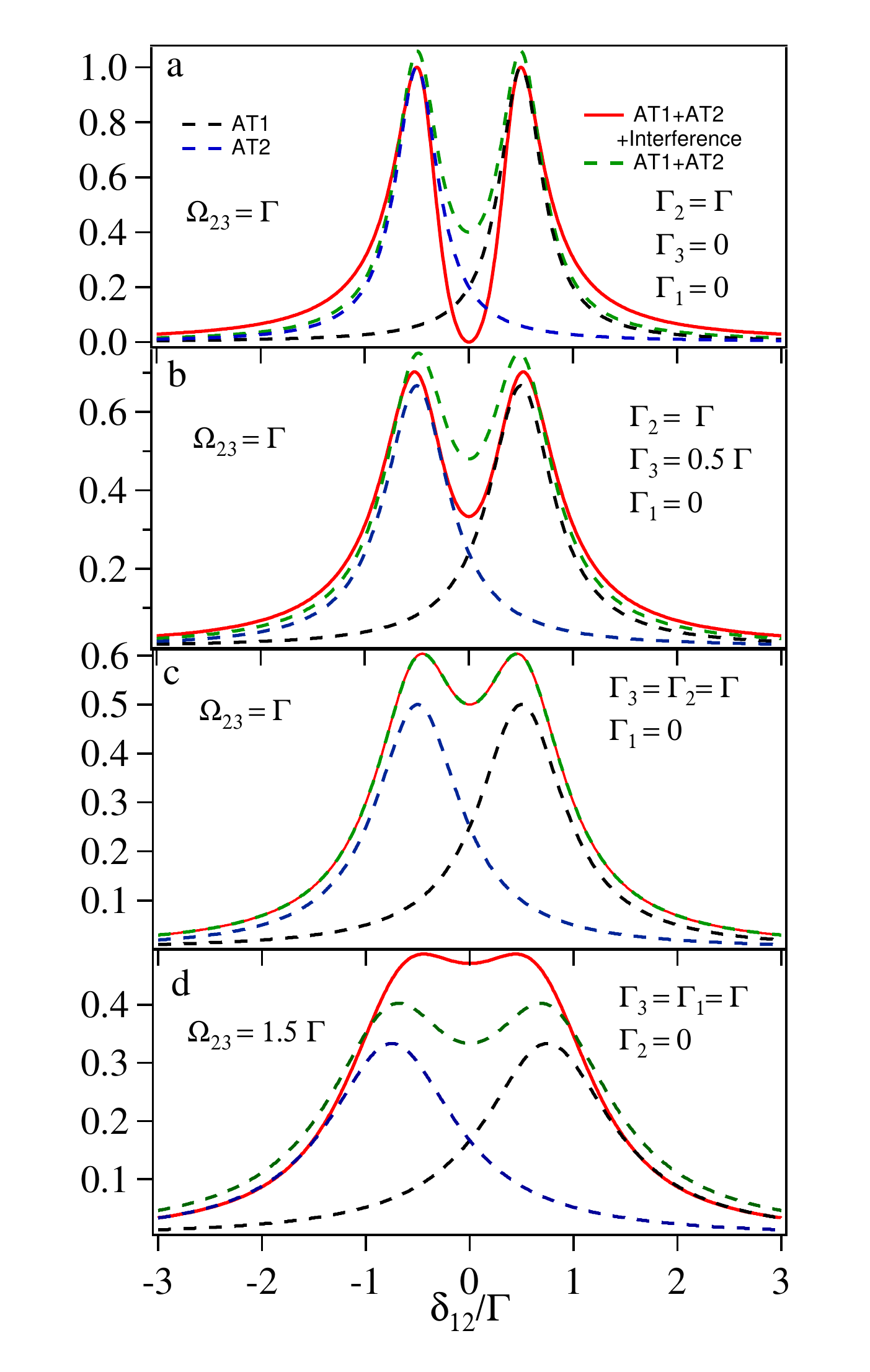}
      \caption {AT peaks and the effect of interference for probe absorption vs  its detuning ($\delta_{12}/\Gamma$) with control laser detuning, $\delta_{23}=0$ i.e. $\theta = \frac{\pi}{4}$ and $\Omega_{p_1}=-\frac{1}{\sqrt{2}}, \Omega_{p_2}=\frac{1}{\sqrt{2}}$  a) $\kappa_{12}=-\frac{\Gamma}{4}, \Gamma_{d_1}=\Gamma_{d_2}= \frac{\Gamma}{4}$  b) $\kappa_{12}=-\frac{\Gamma}{8}, \Gamma_{d_1}= \Gamma_{d_2}=\frac{3\Gamma}{8}$  c) $\kappa_{12}=0, \Gamma_{d_1}=\Gamma_{d_2}=\frac{\Gamma}{2}$  d) $\kappa_{12}=-\frac{\Gamma}{4}, \Gamma_{d_1}= \Gamma_{d_2}=\frac{\Gamma}{4}$. }
      \label{Fig4}
   \end{center}
\end{figure}

From Eq. \ref{eq14}, the interference between the AT peaks is destructive when $\Gamma_2>\Gamma_3$, or no interference when $\Gamma_2=\Gamma_3$, or is constructive when $\Gamma_2<\Gamma_3$ (see also Fig. \ref{Fig4}). For $\Gamma_3=0$ which is the case for a $\Lambda$-system, the destructive interference completely destruct absorption of the two AT peaks at resonance (see Fig. \ref{Fig4}a). In Fig. \ref{Fig4}b with $\Gamma_3=0.5\Gamma_2$ ($\Gamma_2-\Gamma_3>0$) which is valid for the ladder system, the interference is partially destructive as there is a finite absorption for the solid red curve at resonance. For $\Gamma_2=\Gamma_3$ there is no interference between the two AT peaks as the dashed green curve is superimposed on the solid red curve in Fig. \ref{Fig4}c. For $\Gamma_2\ll\Gamma_3$ which is valid for a V-system, the interference is constructive as the dashed green curve is lower than the solid red curve at resonance (see Fig. \ref{Fig4}d). In general, the nature of interference between the two AT peaks can not be tuned from constructive to destructive interference for the given atomic levels by tuning the laser parameters.

Note that in a V-system, the presence of the control laser will cause population transfer. However, in order to study the nature of interference we can ignore population transfer. The population transfer will modify the overall amplitude of absorption but the lineshape will remain the same. The ladder system in Sr using $(5\text{s}^2)^{1}\text{S}_{0},\ket{1}\leftrightarrow(\text{5s5p})^{3}\text{P}_{1},\ket{2}\leftrightarrow(\text{5s6s})^{3}\text{S}_{1},\ket{3}$ with decay parameters $\Gamma_{2}= 2\pi\times7.5$ kHz and $\Gamma_{3}= 2\pi\times16$ MHz is having $\Gamma_{2}\ll\Gamma_{3}$ \cite{CQB05}. This parameter system is very similar to a V-system. In the ladder system, population transfer due to a weak probe laser can be ignored. A similar parameter system is also possible in Yb atom.

\subsection{Four-level system}

\begin{figure}
   \begin{center}
      \includegraphics[width =0.8\linewidth]{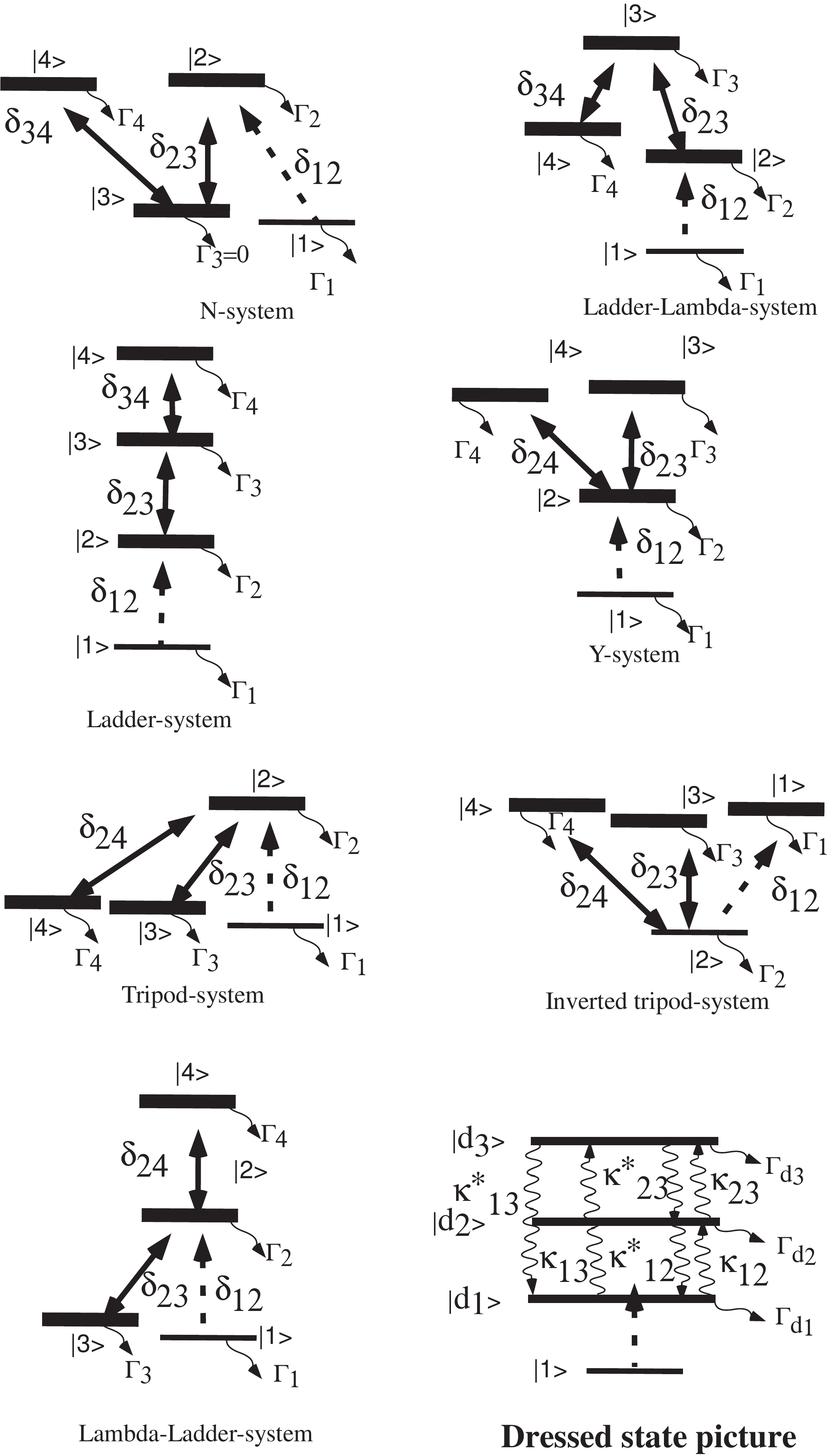}
      \caption{Four-level atomic system in bare and dressed state picture.}
      \label{Fig5}
   \end{center}
\end{figure}

The various four-level system have been studied for different applications in two configurations namely the chain configuration such as N \cite{BMW08,CXH09,SYK11}, Ladder-Lambda \cite{HCN05,JSA12}, Ladder \cite{BJX11}, and branching configuration such as Y \cite{JOX03,GIM18}, Tripod \cite{SLB13,LPR18,HMW14}, Inverted Tripod \cite{PKP16} as shown in Fig. \ref{Fig5}. Different names have been given for the four-level system depending upon the energy levels of $\ket{1}$, $\ket{2}$, $\ket{3}$ and $\ket{4}$ and other possible four-level systems can also be constructed by changing the energy levels. The dressed state picture of all possible four-level systems will be the same as shown in Fig. \ref{Fig5}.

The response of the four level chain system to the weak probe laser in steady state is given by $\rho_{12}$ element of the density matrix using the TOC between levels $\ket{1}$ and $\ket{3}$ and $\ket{1}$ and $\ket{4}$.
\begin{eqnarray}
\label{eq16}
\begin{split}
&\rho_{12}= \cfrac{\frac{i}{2}\frac{\Omega_{12}}{\gamma_{12}}}
{1+\frac{\frac{1}{4}\frac{\mid\Omega_{23}\mid^2}{\gamma_{12}\gamma_{13}}}{1+\frac{1}{4}\frac{\mid\Omega_{34}\mid^2}{\gamma_{13}\gamma_{14}}}}
\end{split}
\end{eqnarray}
For branch system, branching at level $\ket{2}$ the above equation will have the following form
\begin{eqnarray}
\label{eq17}
\begin{split}
&\rho_{12}= \cfrac{\frac{i}{2}\frac{\Omega_{12}}{\gamma_{12}}}
{1+\frac{1}{4}\frac{\mid\Omega_{23}\mid^2}{\gamma_{12}\gamma_{13}}+\frac{1}{4}\frac{\mid\Omega_{24}\mid^2}{\gamma_{12}\gamma_{13}}}
\end{split}
\end{eqnarray}
The above density matrix solution given in Eq. \ref{eq16} and \ref{eq17} is a good cross check of our calculation for the four-level system in the dressed state picture and identification of the nature of interference between the various AT peaks. The Hamiltonian associated with the control lasers for the four-level chain system is written as, 
\begin{equation}
\label{eq18}
H_c=\hbar
\begin{bmatrix}
0 &\frac{\Omega_{23}}{2} &0  \\
\frac{\Omega^*_{23}}{2} &-\delta_{23} & \frac{\Omega_{34}}{2}\\
0 &\frac{\Omega^*_{34}}{2}& -(\delta_{23}\pm\delta_{34}) 
\end{bmatrix}
\end{equation}
Similarly, the Hamiltonian associated with the control lasers for the four-level system branching at level $\ket{2}$ such as Y, Tripod and Inverted Tripod is, 
\begin{equation}
\label{eq19}
H_c=\hbar
\begin{bmatrix}
0 &\frac{\Omega_{23}}{2} &\frac{\Omega_{24}}{2}  \\
\frac{\Omega^*_{23}}{2} &-\delta_{23} & 0\\
\frac{\Omega^*_{24}}{2} &0& -\delta_{34} 
\end{bmatrix}
\end{equation}

The rate equation for the bare state $\ket{1}$ and the dressed states $\ket{d_1}$, $\ket{d_2}$, $\ket{d_3}$ are given below:
\begin{align}
\label{eq20}
i\frac{dC_1}{dt}&=\frac{\Omega_{p_1}}{2}C_{d_1}+\frac{\Omega_{p_2}}{2}C_{d_2}+\frac{\Omega_{p_3}}{2}C_{d_3}\nonumber\\
i\frac{dC_{d_1}}{dt}&=\frac{\Omega^*_{p_1}}{2}C_1-i\gamma_{d_1}C_{d_1}-i\kappa_{12}C_{d_2}-i\kappa_{13}C_{d_3}\\
i\frac{dC_{d_2}}{dt}&=\frac{\Omega^*_{p_2}}{2}C_1-i\kappa^*_{12}C_{d_1}-i\gamma_{d_2}C_{d_2}-i\kappa_{23}C_{d_3}\nonumber\\
i\frac{dC_{d_3}}{dt}&=\frac{\Omega^*_{p_3}}{2}C_1-i\kappa^*_{13}C_{d_1}-i\kappa^*_{23}C_{d_2}-i\gamma_{d_3}C_{d_3}\nonumber
\end{align}
Considering the steady state case for the time evolution of the dressed states i.e. $\frac{dC_{d_1}}{dt}=\frac{dC_{d_2}}{dt}=\frac{dC_{d_3}}{dt}=0$, we get following equation which is proportional to probe absorption.\scriptsize{}
\begin{align}
\label{eq21}
&\frac{dC_1}{dt}=-\frac{1/4}{N}\Bigg[\underbrace{\frac{|\Omega_{p_1}|^2}{\gamma_{d_1}}+\frac{|\Omega_{p_2}|^2}{\gamma_{d_2}}+\frac{|\Omega_{p_3}|^2}{\gamma_{d_3}}}_{\textrm{Three AT peaks}}\nonumber\\
&\underbrace{-\frac{\kappa_{12}\Omega^*_{p_1}\Omega_{p_2}+c.c.}{\gamma_{d_1}\gamma_{d_2}}-\frac{\kappa_{23}\Omega^*_{p_2}\Omega_{p_3}+c.c.}{\gamma_{d_2}\gamma_{d_3}}-\frac{\kappa_{13}\Omega^*_{p_1}\Omega_{p_3}+c.c.}{\gamma_{d_1}\gamma_{d_3}}}_{\textrm{Interference1}}\nonumber\\
&\underbrace{+\frac{\kappa_{12}\kappa_{23}\Omega^*_{p_1}\Omega_{p_3}+c.c.}{\gamma_{d_2}\gamma_{d_1}\gamma_{d_3}}+\frac{\kappa_{23}\kappa^*_{13}\Omega^*_{p_2}\Omega_{p_1}+c.c.}{\gamma_{d_3}\gamma_{d_2}\gamma_{d_1}}+\frac{\kappa^*_{13}\kappa_{12}\Omega^*_{p_3}\Omega_{p_2}+c.c.}{\gamma_{d_1}\gamma_{d_3}\gamma_{d_2}}}_{\textrm{Interference2}}\nonumber\\
&\underbrace{-\frac{|\kappa_{12}|^2|\Omega_{p_3}|^2+|\kappa_{23}|^2|\Omega_{p_1}|^2+|\kappa_{13}|^2|\Omega_{p_2}|^2}{\gamma_{d_1}\gamma_{d_2}\gamma_{d_3}}}_{\textrm{Correction for AT peaks}}\Bigg]
\end{align}
where,
\begin{align}
\label{eq22}
N=1+\frac{[\kappa^*_{13}\kappa_{12}\kappa_{23}+c.c.]}{\gamma_{d_1}\gamma_{d_2}\gamma_{d_3}}-\frac{|\kappa_{12}|^2}{\gamma_{d_1}\gamma_{d_2}}-\frac{|\kappa_{23}|^2}{\gamma_{d_2}\gamma_{d_3}}-\frac{|\kappa_{13}|^2}{\gamma_{d_1}\gamma_{d_3}}
\end{align}

\normalsize{}The above equation represents the absorption of the probe laser. The terms denoted as ``Three AT peaks" represents the absorption of the probe laser due to the three individual dressed states or AT peaks, $\ket{d_1}$, $\ket{d_2}$ and $\ket{d_3}$. The amplitude of the individual AT peaks is proportional to $|\Omega_{p_1}|^2$, $|\Omega_{p_2}|^2$ and $|\Omega_{p_3}|^2$ and is further modified by the ``Correction for AT peaks" term as $-\frac{|\kappa_{23}|^2}{\gamma_{d2}\gamma_{d3}}$, $-\frac{|\kappa_{13}|^2}{\gamma_{d1}\gamma_{d3}}$ and $-\frac{|\kappa_{12}|^2}{\gamma_{d1}\gamma_{d2}}$ respectively. The interference between the AT peaks is denoted by the term ``Interference1" which is very similar to the three-slit interference. The magnitude and sign of the interference is proportional to $\kappa_{12}$, $\kappa_{23}$ and $\kappa_{13}$ respectively.

We also observe the interference between the AT peaks which are little bit more complicated and denoted as ``Interference2". The ``Interference2" is the interference between two AT peaks pair-wise only, but the magnitude and sign are dependent on the coherent decay to the other AT peak. For example, $\kappa_{12}\kappa_{23}\Omega^*_{p_1}\Omega_{p_3}+c.c$ is the interference between $\ket{d_{1}}$ and $\ket{d_{3}}$ but the magnitude and sign is dependent on coherent decay to the other AT peak $\ket{d_{2}}$ through the terms $\kappa_{12}$ and $\kappa_{23}$.

So overall, all the interference terms are pair-wise i.e. like $\Omega^*_{p_1}\Omega_{p_2}+c.c$, $\Omega^*_{p_2}\Omega_{p_3}+c.c$ and $\Omega^*_{p_1}\Omega_{p_3}+c.c$ and not like $\Omega^*_{p_1}\Omega_{p_2}\Omega_{p_3}$. This is again similar to the three-slit interference and it indicates the absence of higher order interference between the Autler-Townes peaks. We have also done similar calculation in five-level system (and other higher levels) and we find the interference to be only pair-wise with no higher order interference.

\subsubsection{Chain configurations with all the control lasers at resonance} 
 
The analytical expression for the eigenvalues and eigenvectors of the Hamiltonian in Eq. \ref{eq18}, are complicated for general detunings. However, when all control lasers are at resonance the expression is simple and easy to interpret the nature of interference between the AT absorption peaks. In this particular case, the eigenvalues of the dressed states and the associated parameters are listed below:
\scriptsize{} 
\begin{align}
\label{eq23}
&E_{d_1}=-\frac{\sqrt{\Omega^2_{23}+\Omega^2_{34}}}{2}; \nonumber\\
&\ket{d_1}=\frac{1}{\sqrt{2}}\frac{\Omega_{23}}{\sqrt{\Omega^2_{23}+\Omega^2_{34}}}\ket{2}-\frac{1}{\sqrt{2}}\ket{3}+\frac{1}{\sqrt{2}}\frac{\Omega_{34}}{\sqrt{\Omega^2_{23}+\Omega^2_{34}}}\ket{4};\nonumber\\
&\Gamma_{d_1}=\frac{1}{4}(\frac{|\Omega_{23}|^2}{\Omega^2_{23}+\Omega^2_{34}}\Gamma_2+\Gamma_3+\frac{|\Omega_{34}|^2}{\Omega^2_{23}+\Omega^2_{34}}\Gamma_4);\nonumber\\
&E_{d_2}=0; \ket{d_2}=-\frac{\Omega_{34}}{\sqrt{\Omega^2_{23}+\Omega^2_{34}}}\ket{2}+\frac{\Omega_{23}}{\sqrt{\Omega^2_{23}+\Omega^2_{34}}}\ket{4};\nonumber\\
&\Gamma_{d_2}=\frac{1}{2}(\frac{|\Omega_{34}|^2}{\Omega^2_{23}+\Omega^2_{34}}\Gamma_2+\frac{|\Omega_{23}|^2}{\Omega^2_{23}+\Omega^2_{34}}\Gamma_4);\nonumber\\
&E_{d_3}=\frac{\sqrt{\Omega^2_{23}+\Omega^2_{34}}}{2}; \nonumber\\
&\ket{d_3}=\frac{1}{\sqrt{2}}\frac{\Omega_{23}}{\sqrt{\Omega^2_{23}+\Omega^2_{34}}}\ket{2}+\frac{1}{\sqrt{2}}\ket{3}+\frac{1}{\sqrt{2}}\frac{\Omega_{34}}{\sqrt{\Omega^2_{23}+\Omega^2_{34}}}\ket{4};\nonumber\\
&\Gamma_{d_3}=\frac{1}{4}(\frac{|\Omega_{23}|^2}{\Omega^2_{23}+\Omega^2_{34}}\Gamma_2+\Gamma_3+\frac{|\Omega_{34}|^2}{\Omega^2_{23}+\Omega^2_{34}}\Gamma_4);\nonumber\\
&\kappa_{12}=-\frac{(\Gamma_2-\Gamma_4)}{2\sqrt{2}}\frac{\Omega_{23}\Omega_{34}}{\Omega^2_{23}+\Omega^2_{34}};\\
&\kappa_{13}=+\frac{\Gamma_2}{4}\frac{\Omega^2_{23}}{\Omega^2_{23}+\Omega^2_{34}}-\frac{\Gamma_3}{4}+\frac{\Gamma_4}{4}\frac{\Omega^2_{34}}{\Omega^2_{23}+\Omega^2_{34}};\nonumber\\
&\kappa_{23}=-\frac{(\Gamma_2-\Gamma_4)}{2\sqrt{2}}\frac{\Omega_{23}\Omega_{34}}{\Omega^2_{23}+\Omega^2_{34}};\nonumber\\
&\Omega_{p_1}=\Omega_{p_3}=\frac{1}{\sqrt{2}}\frac{\Omega_{23}}{\sqrt{\Omega^2_{23}+\Omega^2_{34}}}\Omega_{12};\Omega_{p_2}=-\frac{\Omega_{34}}{\sqrt{\Omega^2_{23}+\Omega^2_{34}}}\Omega_{12}\nonumber
\end{align}
\normalsize{}Various combinations of $\Gamma_2,~\Gamma_3$ and $\Gamma_4$ are considered to see the nature of interference between the AT peaks. 
\begin{enumerate}[(a)]
\item For $\Gamma_2=\Gamma_3=\Gamma_4$: 

The interference parameters $\kappa_{12}=\kappa_{23}=\kappa_{13}=0$ (see Eq. \ref{eq23}), which is implying no interference between any of the AT absorption peaks. The complete overlap of the solid red curve (AT peaks plus interference) and the dashed green curve (AT peaks only) in Fig. \ref{Fig6}a is an indication of no interference.
 
\item For $\Gamma_2<\Gamma_4=\Gamma_3$: 

The interference parameters $\kappa_{12}$ and $\kappa_{23}$ are positive and $\kappa_{13}$ is negative for $\Omega_{23}=\Omega_{34}$ in Eq. \ref{eq23}. The coupling strength $\Omega_{p_1}$ and $\Omega_{p_3}$ are positive and $\Omega_{p_2}$ is negative. Therefore, the overall interference terms of ``Interference1" and ``Interference2" are all positive and the interference between the three AT peaks is constructive. However, the contribution of ``Interference2" and ``Correction for the AT peaks" are negligibly small as shown Fig. \ref{Fig6}b.  

\item For $\Gamma_2>\Gamma_4=\Gamma_3$: 

The interference parameters $\kappa_{12}$ and $\kappa_{23}$ are negative and $\kappa_{13}$ is positive for $\Omega_{23}=\Omega_{34}$ in Eq. \ref{eq23}. Similarly, $\Omega_{p_1}$ and $\Omega_{p_3}$ are positive while $\Omega_{p_2}$ is negative. This implies that all the interference terms in ``Interference1" are negative and hence there is destructive interference between the three AT peaks. The destructive interference however, partially reduces absorption of the AT peaks at the region of overlap. The interference terms of ``Interference2" are positive and the contribution of the constructive interference is very small. The correction to individual AT peaks is equally negligible as shown in Fig. \ref{Fig6}c. 

\item For $\Gamma_2>\Gamma_4=\Gamma_3=0$: 

The interference parameters $\kappa_{12}$ and $\kappa_{23}$ are negative and $\kappa_{13}$ is positive. Hence, all the terms in ``Interference1" are negative and in ``Interference2" are positive. The destructive interference completely destruct absorption of the three AT peaks at the overlap region giving rise to a double transparency window (see Fig. \ref{Fig6}d). The correction to the individual AT peaks from the terms in ``Correction for AT peaks" is significant while the interference contribution of ``Interference2" is small.

\begin{figure}
   \begin{center}
      \includegraphics[width =0.8\linewidth]{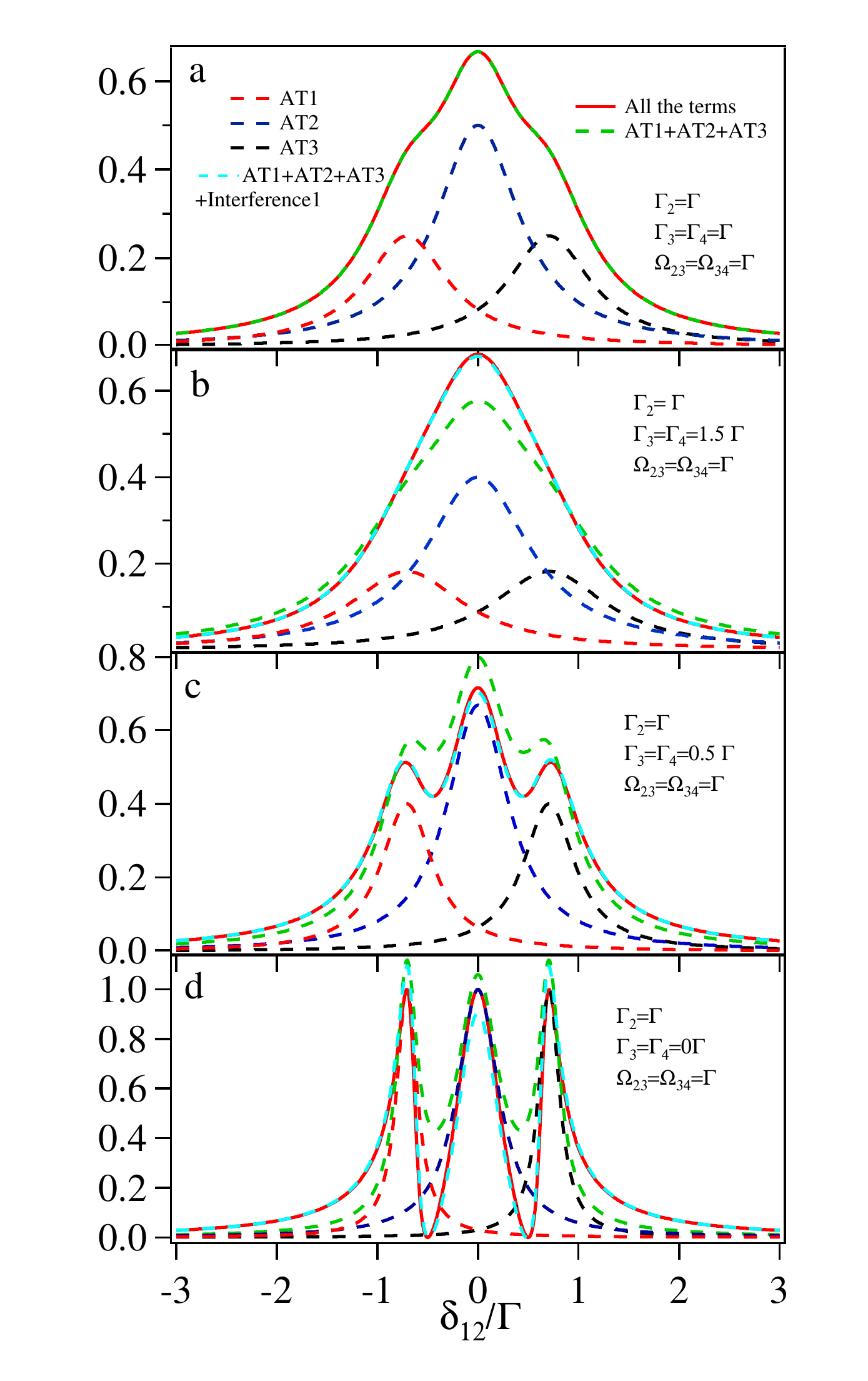}
      \caption{AT peaks and the effect of interference for probe absorption vs its detuning ($\delta_{12}/\Gamma$) with $\delta_{23}=\delta_{34}=0$ and $\Omega_{23}=\Omega_{34}$ i.e. $\Omega_{p_1}=\Omega_{p_3}=\frac{\Omega_{12}}{2}$, $\Omega_{p_2}=-\frac{\Omega_{12}}{\sqrt{2}}$. a) $\kappa_{12}=\kappa_{23}=\kappa_{13}=0$, $\Gamma_{d_1}=\Gamma_{d_2}=\Gamma_{d_3}=\frac{\Gamma}{2}$, b) $\kappa_{12}=\kappa_{23}=+\frac{\Gamma_2}{8\sqrt{2}}$, $\kappa_{13}=-\frac{\Gamma_2}{16}$, $\Gamma_{d_1}=\Gamma_{d_3}=\frac{11\Gamma}{16}$, $\Gamma_{d_2}=\frac{5\Gamma}{8}$ c) $\kappa_{12}=\kappa_{23}=-\frac{\Gamma_2}{8\sqrt{2}}$, $\kappa_{13}=+\frac{\Gamma_2}{16}$, $\Gamma_{d_1}=\Gamma_{d_3}=\frac{5\Gamma}{16}$, $\Gamma_{d_2}=\frac{3\Gamma}{8}$, d) $\kappa_{12}=\kappa_{23}=-\frac{\Gamma_2}{4\sqrt{2}}$, $\kappa_{13}=+\frac{\Gamma_2}{8}$, $\Gamma_{d_1}=\Gamma_{d_3}=\frac{\Gamma}{8}$, $\Gamma_{d_2}=\frac{\Gamma}{4}$.}
      \label{Fig6}
   \end{center}
\end{figure}
\begin{figure}
   \begin{center}
      \includegraphics[width =0.8\linewidth]{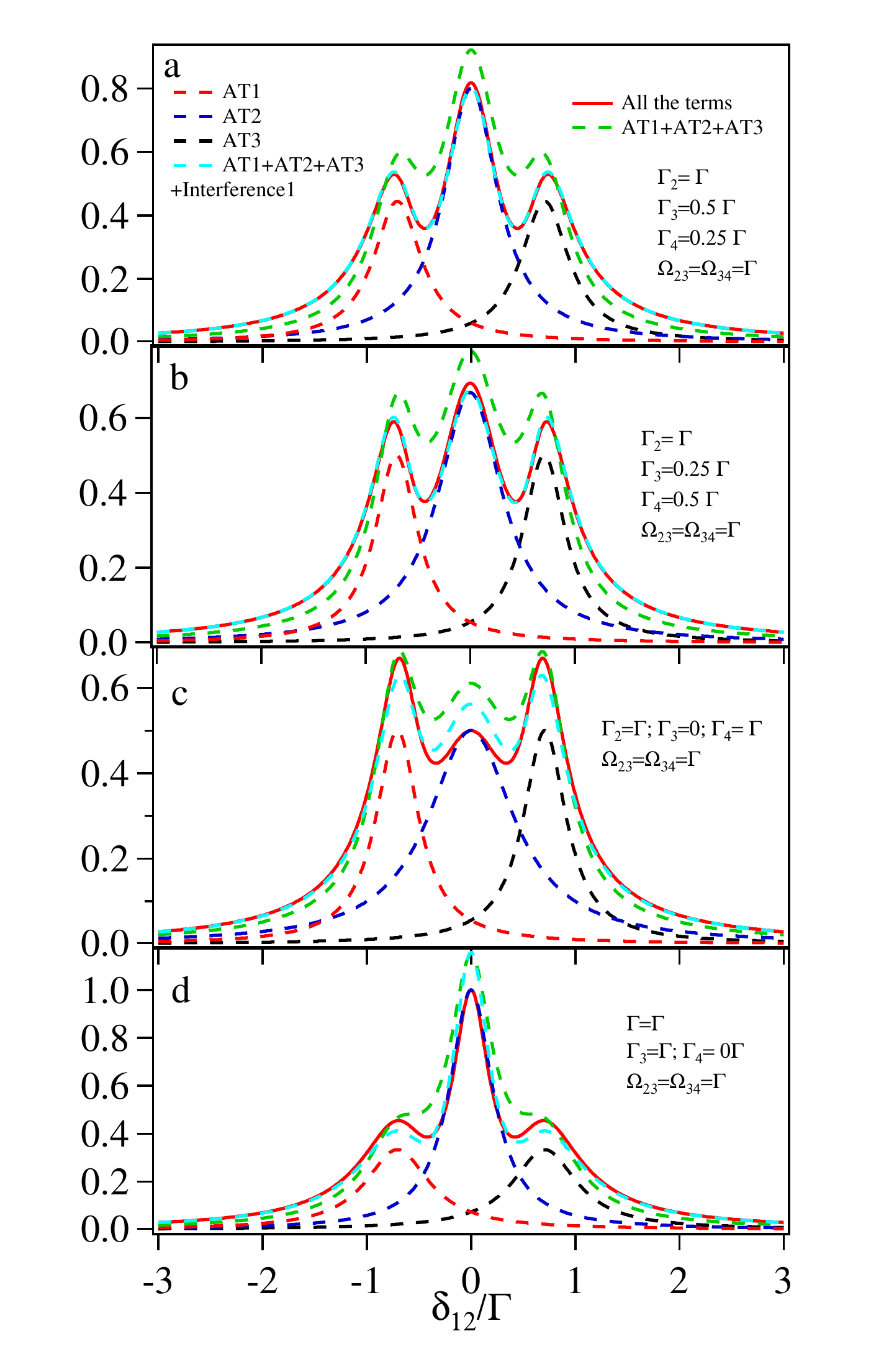}
      \caption{AT peaks and the effect of interference for probe absorption vs  its detuning ($\delta_{12}/\Gamma$) with $\delta_{23}=\delta_{34}=0$ and $\Omega_{23}=\Omega_{34}$ i.e. $\Omega_{p_1}=\Omega_{p_3}=\frac{\Omega_{12}}{2}$, $\Omega_{p_2}=-\frac{\Omega_{12}}{\sqrt{2}}$. a) $\kappa_{12}=\kappa_{23}=-\frac{3\Gamma_2}{16\sqrt{2}}$, $\kappa_{13}=+\frac{\Gamma_2}{32}$, $\Gamma_{d_1}=\Gamma_{d_3}=\frac{9\Gamma}{32}$, $\Gamma_{d_2}=\frac{5\Gamma}{16}$ b) $\kappa_{12}=\kappa_{23}=-\frac{\Gamma_2}{8\sqrt{2}}$, $\kappa_{13}=+\frac{\Gamma_2}{8}$, $\Gamma_{d_1}=\Gamma_{d_3}=\frac{\Gamma}{4}$, $\Gamma_{d_2}=\frac{3\Gamma}{8}$ c) $\kappa_{12}=\kappa_{23}=0$, $\kappa_{13}=+\frac{\Gamma_2}{4}$, $\Gamma_{d_1}=\Gamma_{d_3}=\frac{\Gamma}{4}$, $\Gamma_{d_2}=\frac{\Gamma}{2}$ d) $\kappa_{12}=\kappa_{23}=-\frac{\Gamma_2}{4\sqrt{2}}$, $\kappa_{13}=-\frac{\Gamma_2}{8}$, $\Gamma_{d_1}=\Gamma_{d_3}=\frac{3\Gamma}{8}$, $\Gamma_{d_2}=\frac{\Gamma}{2}$.}
      \label{Fig7}
   \end{center}
\end{figure}
\item For $\Gamma_2  > \Gamma_4 \neq \Gamma_3$: 

The interference parameters $\kappa_{12}$ and $\kappa_{23}$ are negative and $\kappa_{13}$ is positive for $\Omega_{23}=\Omega_{34}$ in Fig. \ref{Fig7}a and \ref{Fig7}b and hence, all the terms in``Interference1" are negative and in ``Interference2" are positive. However, the contribution from ``Interference2" is very small and the interference between the AT peaks is essentially destructive. The modification of the individual AT peaks by ``Correction for AT peaks" terms is also negligible.
 
\item For $\Gamma_3=0, \Gamma_4 = \Gamma_2$ (N-system):

Both the interference parameters $\kappa_{12}$ and $\kappa_{23}$ are zero but $\kappa_{13}$ is positive. This implies that the first two terms in ``Interference1" are zero and the third term is negative. All the terms in ``Interference2" are zero and therefore the overall interference between $\ket{d_1}$ and $\ket{d_3}$ is destructive which partially reduce the absorption of the AT peaks. The contribution of ``Correction for AT peaks" is significant as seen by the deviation of the dashed cyan curve from the solid red curve (see Fig. \ref{Fig7}c).
 
\item For $\Gamma_3=\Gamma_2, \Gamma_4=0$ (Ladder-lambda system):

The interference parameters $\kappa_{12}$, $\kappa_{23}$  and $\kappa_{13}$ are all negative (see Fig. \ref{Fig7}d). Hence, the first two terms in ``Interference1" are negative and the third term is positive. Similarly, the first two terms in ``Interference2" are negative and the third term is positive. In the said figure destructive interference between the three AT peaks dominates and both ``Interference2" and ``Correction for AT peaks" have a significant contribution.
\end{enumerate} 

\begin{figure}
   \begin{center}      
      \includegraphics[width =0.8\linewidth]{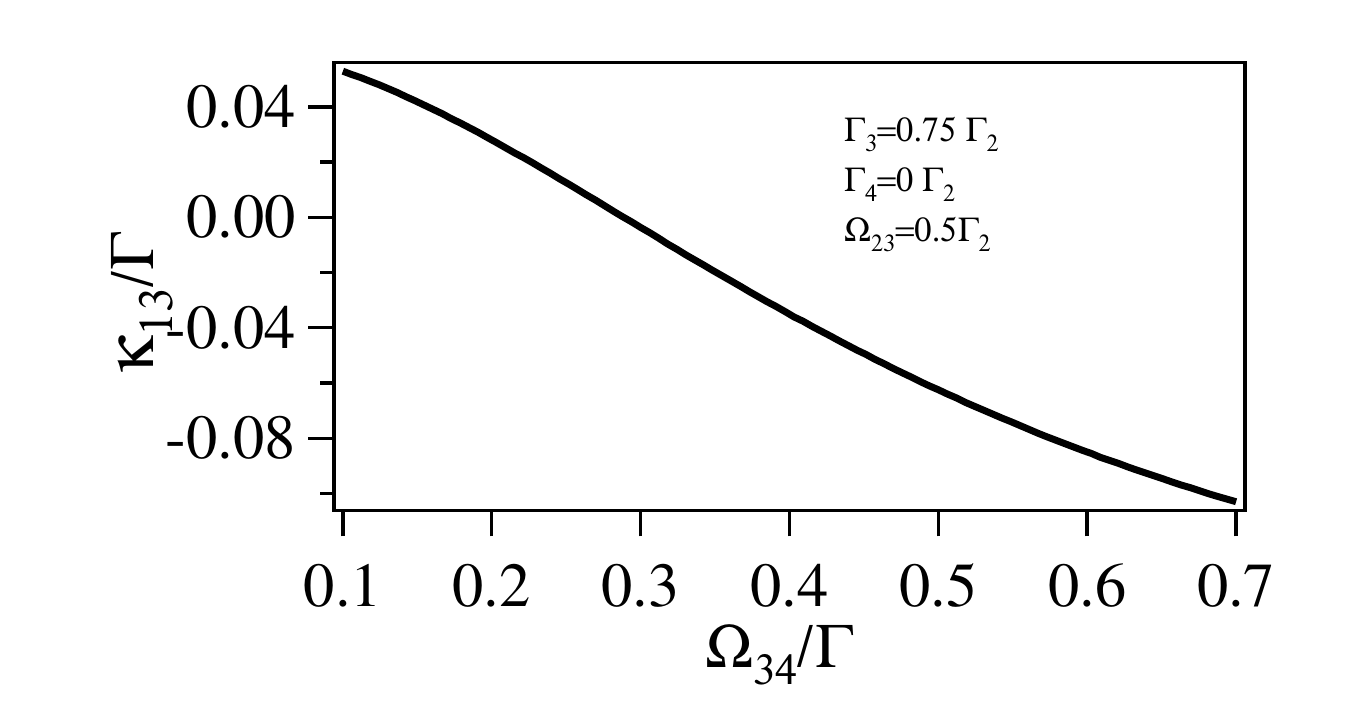} 
      \caption{Variation of coherent decay between dressed state $\ket{d_1}$ and $\ket{d_3}$, i.e. $\kappa_{13}$ as a function of the coupling strength $\Omega_{34}$ of the control laser.}
      \label{Fig8}
   \end{center}
\end{figure}

The possibility of tuning the interference from constructive to destructive by tuning the control laser parameter such as the Rabi frequencies is illustrated in Fig. \ref{Fig8}. In the three-level system we noted that the nature of interference can not be tuned from negative to positive or zero by changing the laser parameters. However, if we consider a simple case in four-level-system where $\Gamma_3<\Gamma_2$ and $\Gamma_4=0$ it is possible to tune the interference from negative to positive between $\ket{d_1}$ and $\ket{d_3}$. In Fig. \ref{Fig8} we plot the interference parameter $\kappa_{13}$ vs $\Omega_{34}$  for a system with $\Gamma_3=0.75\Gamma_2$ and $\Gamma_4=0$. The interference parameter $\kappa_{13}$ is positive, zero and negative for low, $1/\sqrt{12}\Gamma_2$ and high value of $\Omega_{34}$ respectively. Also as the value of $\Omega_{34}$ increases, there is no change of sign in $\Omega_{p_1}$, $\Omega_{p_2}$ and $\Omega_{p_3}$. Therefore, it is possible to tune the interference between $\ket{d_1}$ and $\ket{d_3}$ from constructive, to zero and to destructive.

\subsubsection{Branching configurations with all control lasers at resonance}

We similarly consider the case when all the control lasers are at resonance for the Hamiltonian given in Eq. \ref{eq19}. In this particular case the energy of the dressed states and the related parameters are listed below: 
\scriptsize{} 
\begin{align}
\label{eq24}
&E_{d_1}=-\frac{\sqrt{\Omega^2_{23}+\Omega^2_{24}}}{2};\nonumber\\
&\ket{d_1}=\frac{1}{\sqrt{2}}\frac{\Omega_{23}}{\sqrt{\Omega^2_{23}+\Omega^2_{24}}}\ket{3}-\frac{1}{\sqrt{2}}\ket{2}+\frac{1}{\sqrt{2}}\frac{\Omega_{24}}{\sqrt{\Omega^2_{23}+\Omega^2_{24}}}\ket{4};\nonumber\\
&\Gamma_{d1}=\frac{1}{4}(\frac{|\Omega_{23}|^2}{\Omega^2_{23}+\Omega^2_{24}}\Gamma_3+\Gamma_2+\frac{|\Omega_{24}|^2}{\Omega^2_{23}+\Omega^2_{24}}\Gamma_4);\nonumber\\
&E_{d_2}=0; \ket{d_2}=-\frac{\Omega_{24}}{\sqrt{\Omega^2_{23}+\Omega^2_{24}}}\ket{3}+\frac{\Omega_{23}}{\sqrt{\Omega^2_{23}+\Omega^2_{24}}}\ket{4};\nonumber\\
&\Gamma_{d_2}=\frac{1}{2}(\frac{|\Omega_{24}|^2}{\Omega^2_{23}+\Omega^2_{24}}\Gamma_3+\frac{|\Omega_{23}|^2}{\Omega^2_{23}+\Omega^2_{24}}\Gamma_4);\nonumber\\
&E_{d_3}=\frac{\sqrt{\Omega^2_{23}+\Omega^2_{24}}}{2};\nonumber\\
&\ket{d_3}=\frac{1}{\sqrt{2}}\frac{\Omega_{23}}{\sqrt{\Omega^2_{23}+\Omega^2_{24}}}\ket{3}+\frac{1}{\sqrt{2}}\ket{2}+\frac{1}{\sqrt{2}}\frac{\Omega_{24}}{\sqrt{\Omega^2_{23}+\Omega^2_{24}}}\ket{4};\nonumber\\
&\Gamma_{d_3}=\frac{1}{4}(\frac{|\Omega_{23}|^2}{\Omega^2_{23}+\Omega^2_{24}}\Gamma_3+\Gamma_2+\frac{|\Omega_{24}|^2}{\Omega^2_{23}+\Omega^2_{24}}\Gamma_4);\nonumber\\
&\kappa_{12}=-\frac{(\Gamma_3-\Gamma_4)}{2\sqrt{2}}\frac{\Omega_{23}\Omega_{24}}{\Omega^2_{23}+\Omega^2_{24}};\\
&\kappa_{13}=\frac{\Gamma_3}{4}\frac{\Omega^2_{23}}{\Omega^2_{23}+\Omega^2_{24}}-\frac{\Gamma_2}{4}+\frac{\Gamma_4}{4}\frac{\Omega^{2}_{24}}{\Omega^2_{23}+\Omega^2_{24}};\nonumber\\
&\kappa_{23}=-\frac{(\Gamma_3-\Gamma_4)}{2\sqrt{2}}\frac{\Omega_{23}\Omega_{24}}{\Omega^2_{23}+\Omega^2_{24}}; \Omega_{p_1}=-\Omega_{p_3}=\frac{\Omega_{12}}{\sqrt{2}};  \Omega_{p_2}=0;\nonumber
\end{align}
\normalsize{}
\begin{figure}
   \begin{center}
      \includegraphics[width =0.8\linewidth]{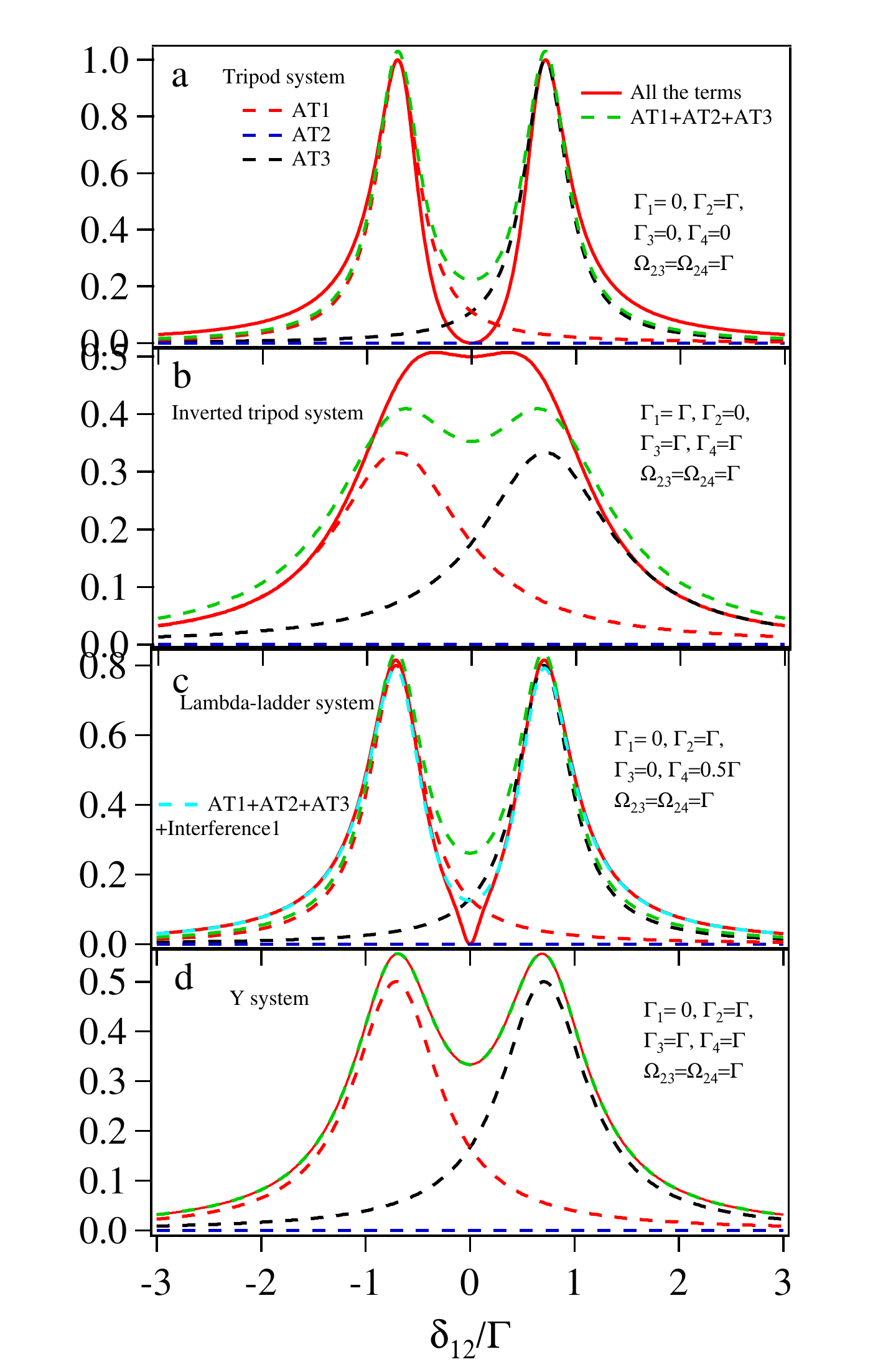}
      \caption{AT peaks and the effect of interference for probe absorption vs  its detuning ($\delta_{12}/\Gamma$) with $\delta_{23}=\delta_{24}=0$ and $\Omega_{23}=\Omega_{24}$ i.e. $\Omega_{p_1}=-\Omega_{p_3}=\frac{\Omega_{12}}{\sqrt{2}}$, $\Omega_{p_2}=0$ a) Tripod-system, $\kappa_{12}=\kappa_{23}=0$, $\kappa_{13}=-\frac{\Gamma}{4}$, $\Gamma_{d_1}=\Gamma_{d_3}=\frac{\Gamma}{4}$, $\Gamma_{d_2}=0$  b) Inverted tripod system $\kappa_{12}=\kappa_{23}=0$, $\kappa_{13}=+\frac{\Gamma}{4}$, $\Gamma_{d_1}=\Gamma_{d_3}=\frac{\Gamma}{4}$, $\Gamma_{d_2}=\frac{\Gamma}{2}$ c) Lambda-ladder-system $\kappa_{12}=\kappa_{23}=+\frac{\Gamma}{8\sqrt{2}}$, $\kappa_{13}=\frac{3\Gamma}{16}$, $\Gamma_{d_1}=\Gamma_{d_3}=\frac{5\Gamma}{16}$, $\Gamma_{d_2}=\frac{\Gamma}{8}$ d) $\kappa_{12}=\kappa_{23}=\kappa_{13}=0$, $\Gamma_{d_1}=\Gamma_{d_2}=\Gamma_{d_3}=\frac{\Gamma}{2}$.}
      \label{Fig9}
   \end{center}
\end{figure}
The central AT peak shown by dashed blue curve (corresponding to eigenvalue 0) in Fig. \ref{Fig9} has zero amplitude as $\Omega_{p_2}=0$. Therefore, only two dressed states $\ket{d_1}$ and $\ket{d_3}$ can be excited by the probe laser and the branching system behaves like the three-level system. 
\begin{enumerate}[(a)]
\item For $\Gamma_3=\Gamma_4=0$ (Tripod system \cite{LPR18}):

 All the terms in ``Interference2" and ``Correction for AT peaks" are zero. The destructive interference between $\ket{d_1}$ and $\ket{d_3}$ (since the ``Interference1" term $-\kappa_{13}\Omega_{p_1}\Omega_{p_3}$ is negative) completely destructs absorption of the two AT peaks at resonance as seen in Fig. \ref{Fig9}a. 

\item For $\Gamma_2=0$, $\Gamma_1=\Gamma_3=\Gamma_4=\Gamma$ (Inverted tripod system \cite{PKP16}): 

All the terms in ``Interference2" and ``Correction for AT peaks" are also zeros and the constructive interference between $\ket{d_1}$ and $\ket{d_3}$ (since the ``Interference1" term -$\kappa_{13}\Omega_{p_1}\Omega_{p_3}$ is positive) enhance absorption (see Fig. \ref{Fig9}b). 

\item For $\Gamma_1=0$, $\Gamma_2=\Gamma$, $\Gamma_3=0$ and $\Gamma_4=0.5\Gamma$ (Ladder-lambda system):

The ``Interference1" term -$\kappa_{13}\Omega_{p_1}\Omega_{p_3}$ between $\ket{d_1}$ and $\ket{d_3}$ is negative and hence there is destructive interference. The non-zero term of ``Interference2" $\kappa_{12}\kappa_{23}\Omega^*_{p_1}\Omega_{p_3}$ is negative and contribute to destructive interference. The ``Correction for the AT peaks " is also non-zero for the two dressed states leading to further reduction in the absorption as shown in Fig. \ref{Fig9}c. 

\item For $\Gamma_2=\Gamma_3=\Gamma_4=\Gamma$ which is valid for (Y-system):

The terms ``Interference1", ``Interference2" and ``Correction for AT peaks" are zero as $\kappa_{12}=\kappa_{13}=\kappa_{23}=0$. There is no interference between any of the AT peaks as evidenced by the complete overlap between the solid red curve and the green dash curve in Fig. \ref{Fig9}d.       
\end{enumerate} 

\subsection{Four-level loopy system}

\begin{figure}
   \begin{center}
      \includegraphics[width =0.8\linewidth]{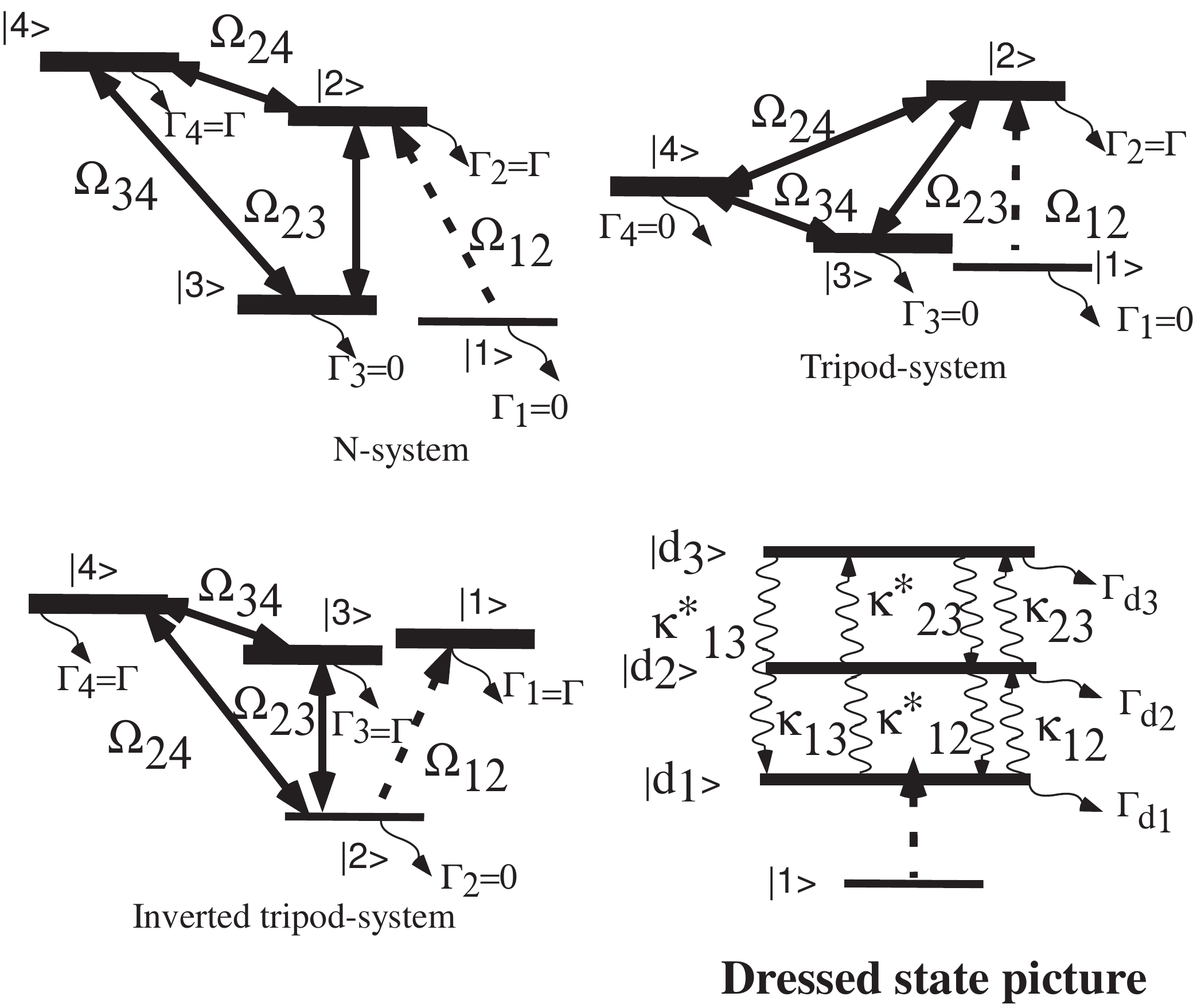}
      \caption{Four-level loopy system in bare and dressed state picture.}
      \label{Fig10}
   \end{center}
\end{figure}

The four-level loopy system is as shown in Fig. \ref{Fig10}. It can be of various types depending upon the energy level of the states $\ket{1}$, $\ket{2}$, $\ket{3}$, $\ket{4}$  as we discussed in the previous section. The study of the loopy system is very important as it further authenticates our approach for the dressed states as in this case the interference terms, $\kappa's$ and $\Omega_{p}'s$ can be complex. The various loopy system has been discussed previously \cite{MDP17,SOP18,SHP18,TDL18}. 
The density matrix element for the probe absorption in the four-level loopy system shown in Fig. \ref{Fig10} is given by the following equation, 
\begin{eqnarray}
\label{eq25}
\begin{split}
&\rho_{12}= \cfrac{\frac{i}{2}\frac{\Omega_{12}}{\gamma_{12}}}
{1+\cfrac{\frac{1}{4}\frac{\mid\Omega_{23}\mid^2}{\gamma_{12}\gamma_{13}}+\frac{1}{4}\frac{\mid\Omega_{24}\mid^2}{\gamma_{12}\gamma_{14}}+\frac{i}{8}\frac{\Omega_{23}^{\ast}\Omega_{34}^{\ast}\Omega_{24}+c.c.}{\gamma_{12}\gamma_{13}\gamma_{14}}}{1+\frac{1}{4}\frac{\mid\Omega_{34}\mid^2}{\gamma_{13}\gamma_{14}}}}
\end{split}
\end{eqnarray}

The Hamiltonian associated with the control lasers for this system is given below
\begin{equation}
\label{eq26}
H_c=\hbar
\begin{bmatrix}
0 &\frac{\Omega_{23}}{2} &\frac{\Omega_{24}}{2}  \\
\frac{\Omega^*_{23}}{2} &-\delta_{23} & \frac{\Omega_{34}}{2}\\
\frac{\Omega^*_{24}}{2} &\frac{\Omega^*_{34}}{2}& -\delta_{34} 
\end{bmatrix}
\end{equation}
The general control laser Rabi frequencies will be $\Omega_{23}=|\Omega_{23}|e^{i\phi_{23}}$, $\Omega_{34}=|\Omega_{34}|e^{i\phi_{34}}$ and $\Omega_{24}=|\Omega_{24}|e^{i\phi}$ which can be considered (without loss of generality) as $\Omega_{23}=|\Omega_{23}|$ and $\Omega_{34}=|\Omega_{34}|$ and $\Omega_{24}=|\Omega_{24}|e^{i\phi}$ i.e. considering $\Omega_{23}$ and $\Omega_{34}$ as real and $\Omega_{24}=|\Omega_{24}|e^{i\phi}$ as complex where $\phi=\phi_{24}-\phi_{23}-\phi_{34}$. The eigenvalue and eigenvector corresponding to general parameters in the above Hamiltonian given in Eq. \ref{eq26} is complicated. However, it is relatively simpler for the case where all the control lasers are at resonance and $\Omega_{23}=\Omega_{34}=\Omega$ and $\Omega_{24}=\Omega_1e^{i\phi}$ where $\Omega$ and $\Omega_1$ are real quantities.

\begin{figure}
   \begin{center}
      \includegraphics[width =0.8\linewidth]{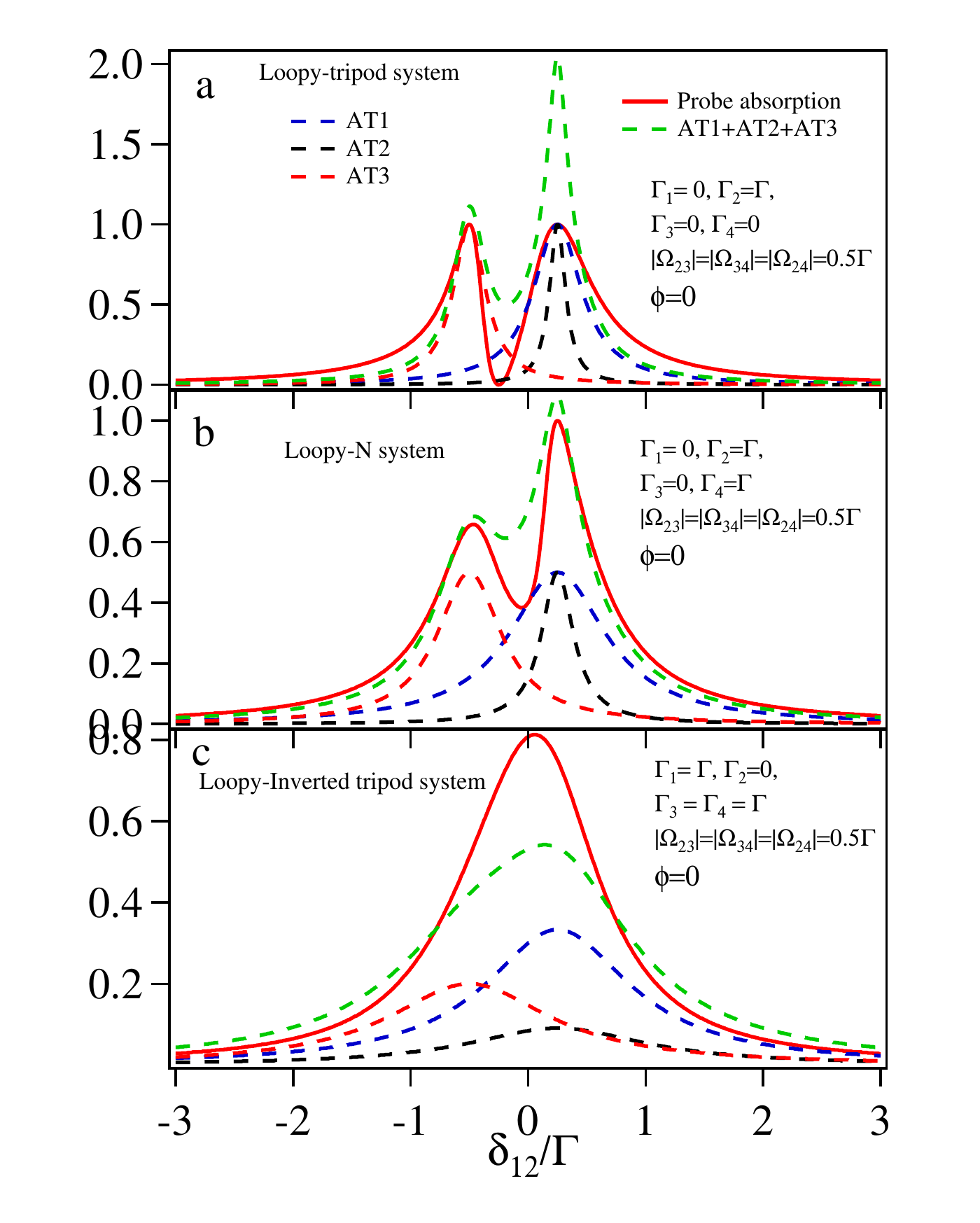}
        \caption{AT peaks and the effect of interference for probe absorption vs its detuning ($\delta_{12}/\Gamma$) with $\delta_{23}=\delta_{34}=\delta_{24}=0$, $\Omega_{23}=\Omega_{34}=\Omega_{24}$ i.e. $\Omega_{p_1}=-\frac{\Omega_{12}}{\sqrt{2}}$, $\Omega_{p_2}=\frac{\Omega_{12}}{\sqrt{6}}$, $\Omega_{p_3}=\frac{\Omega_{12}}{\sqrt{3}}$ and $\phi=0$. a) $\kappa_{12}=-\frac{\Gamma}{4\sqrt{3}}$, $\kappa_{23}=+\frac{\Gamma}{6\sqrt{2}}$, $\kappa_{13}=-\frac{\Gamma}{2\sqrt{6}}$, $\Gamma_{d_1}=\frac{\Gamma}{4}$, $\Gamma_{d_2}=\frac{\Gamma}{12}$ and $\Gamma_{d_3}=\frac{\Gamma}{6}$ b) $\kappa_{12}=\kappa_{13}=0$, $\kappa_{23}=+\frac{\Gamma}{3\sqrt{2}}$, $\Gamma_{d_1}=\frac{\Gamma}{2}$, $\Gamma_{d_2}=\frac{\Gamma}{6}$ and $\Gamma_{d_3}=\frac{\Gamma}{3}$ c) $\kappa_{12}=+\frac{\Gamma}{4\sqrt{3}}$, $\kappa_{23}=-\frac{\Gamma}{6\sqrt{2}}$, $\kappa_{13}=+\frac{\Gamma}{2\sqrt{6}}$, $\Gamma_{d_1}=\frac{\Gamma}{4}$, $\Gamma_{d_2}=\frac{5\Gamma}{12}$ and $\Gamma_{d_3}=\frac{\Gamma}{3}$.}
      \label{Fig11}
   \end{center}
\end{figure}

The eigenvalues, eigenvectors and the various other parameters for $\phi=0$ are listed below:
\scriptsize{}
\begin{align}
\label{eq27}
&E_{d_1}=-\frac{\Omega_1}{2}; \ket{d_1}=-\sqrt{\frac{1}{2}}\ket{2}+\sqrt{\frac{1}{2}}\ket{4}; \Gamma_{d_1}=\frac{\Gamma_2}{4}+\frac{\Gamma_4}{4}; \nonumber\\
&E_{d_2}=\frac{\Omega_1-\Omega^{\prime\prime}}{4};\ket{d_2}=\frac{\sqrt{2}\Omega\Big[\ket{2}-\Big(\frac{ \Omega_1+\Omega^{\prime\prime}}{2\Omega}\Big)\ket{3}+\ket{4}\Big]}{\sqrt{\Omega^{\prime\prime}(\Omega^{\prime\prime}+\Omega_1)}}\nonumber\\
&\Gamma_{d_2}=\frac{\Omega^2\Big[\Gamma_2+\Big(\frac{ \Omega_1+\Omega^{\prime\prime}}{2\Omega}\Big)^2\Gamma_3+\Gamma_4\Big]}{\Omega^{\prime\prime}(\Omega^{\prime\prime}+\Omega_1)}\nonumber\\
&E_{d_3}=\frac{\Omega_1+\Omega^{\prime\prime}}{4}; \ket{d_3}=\frac{\sqrt{2}\Omega\Big[\ket{2}+\Big(\frac{\Omega^{\prime\prime}-\Omega_1}{2\Omega}\Big)\ket{3}+\ket{4}\Big]}{\sqrt{\Omega^{\prime\prime}(\Omega^{\prime\prime}-\Omega_1)}}\nonumber\\
&\Gamma_{d_3}=\frac{\Omega^2\Big[\Gamma_2+\Big(\frac{\Omega^{\prime\prime}-\Omega_1}{2\Omega}\Big)^2\Gamma_3+\Gamma_4\Big]}{\Omega^{\prime\prime}(\Omega^{\prime\prime}-\Omega_1)}\\
&\kappa_{12}=-\frac{(\Gamma_2-\Gamma_4)\Omega}{2\sqrt{\Omega^{\prime\prime}(\Omega^{\prime\prime}+\Omega_1)}};~\kappa_{23}=\frac{(+\Gamma_2-2\Gamma_3+\Gamma_4)\Omega}{2\sqrt{2}\Omega^{\prime\prime}}\nonumber\\
&\kappa_{13}=-\frac{(\Gamma_2-\Gamma_4)\Omega}{2\sqrt{\Omega^{\prime\prime}(\Omega^{\prime\prime}-\Omega_1)}};~\Omega_{p_1}=-\frac{\Omega_{12}}{\sqrt{2}}; \Omega_{p_2}=\frac{\sqrt{2}\Omega}{\sqrt{\Omega^{\prime\prime}(\Omega^{\prime\prime}+\Omega_1)}}\Omega_{12}\nonumber\\
&\Omega_{p_3}=\frac{\sqrt{2}\Omega}{\sqrt{\Omega^{\prime\prime}(\Omega^{\prime\prime}-\Omega_1)}}\Omega_{12};~\textrm{where,} \Omega^{\prime\prime}=\sqrt{8\Omega^2+\Omega_1^2}\nonumber
\end{align}
\normalsize{}
We consider the various combinations of $\Gamma_2,~\Gamma_3$ and $\Gamma_4$ to see the nature of interference between the AT peaks and take $\Omega_{23}=\Omega_{34}=\Omega_{24}=0.5\Gamma$ for all the cases of the loopy systems for $\phi=0$. 
\begin{enumerate}[(i)]

\item For $\Gamma_1=\Gamma_3=\Gamma_4=0$, $\Gamma_2=\Gamma$ (loopy-tripod-system):

All the individual terms of ``Interference1" are negative and  of ``Interference2" are positive but negligibly small. The destructive interference completely destructs probe absorption at the crossing regions of the AT peaks as shown in Fig. \ref{Fig11}a. 
 
\item For $\Gamma_2=\Gamma_4=\Gamma$, $\Gamma_1=\Gamma_3=0$ (loopy N-system):

The interference terms $\kappa_{12}=\kappa_{13}=0$ and $\kappa_{23}$ is negative. Hence, corresponding to ``Interference1" there is no interference between the AT peaks $\ket{d_1}$ and $\ket{d_2}$ and $\ket{d_1}$ and $\ket{d_3}$ while there is destructive interference between $\ket{d_2}$ and $\ket{d_3}$. For this system all the terms corresponding to ``Interference2" are zero. The probe absorption and the AT peaks for this particular system are shown in Fig. \ref{Fig11}b. 

\item For $\Gamma_1=\Gamma_3=\Gamma_4=\Gamma$, $\Gamma_2=0$ (loopy inverted-tripod system):

All the individual terms of the ``Interference1" and ``Interference2" are positive and hence there is constructive interference between the AT peaks (see Fig. \ref{Fig11}c).
\end{enumerate}

For $\phi=\pi/2$ the different parameters for the dressed states are given below:
\scriptsize{}
\begin{align}
\label{eq28}
&E_{d_1}=-\frac{\Omega'}{2}; \ket{d_1}=\frac{\sqrt{\Omega_1^2+\Omega^2}}{\sqrt{2}\Omega'}\Bigg[\frac{\Omega'-i\Omega_1}{\Omega'+i\Omega_1}\ket{2}-\frac{2\Omega}{\Omega'+i\Omega_1}\ket{3}+\ket{4}\Bigg]\nonumber\\
&\Gamma_{d_1}=\frac{\Omega_1^2+\Omega^2}{4\Omega'^2}\Bigg[\Gamma_2+\frac{4\Omega^2}{\Omega'^2+\Omega^2}\Gamma_3+\Gamma_4\Bigg]\nonumber\\
&E_{d_2}=0; \ket{d_2}=-\frac{\Omega}{\Omega'}\ket{2}-\frac{i\Omega_1}{\Omega'}\ket{3}+\frac{\Omega}{\Omega'}\ket{4}\nonumber\\
& \Gamma_{d_2}=\frac{1}{2}(\frac{\Omega^2}{\Omega'^2}\Gamma_2+\frac{\Omega_1^2}{\Omega'^2}\Gamma_3+\frac{\Omega^2}{\Omega'^2}\Gamma_4)\nonumber\\
&E_{d_3}=\frac{\Omega'}{2}; \ket{d_3}=\frac{\sqrt{\Omega_1^2+\Omega^2}}{\sqrt{2}\Omega'}\Bigg[\frac{\Omega'+i\Omega_1}{\Omega'-i\Omega_1}\ket{2}+\frac{2\Omega}{\Omega'-i\Omega_1}\ket{3}+\ket{4}\Bigg]\nonumber\\
&\Gamma_{d_3}=\frac{\Omega_1^2+\Omega^2}{4\Omega'^2}\Bigg[\Gamma_2+\frac{4\Omega^2}{\Omega'^2+\Omega^2}\Gamma_3+\Gamma_4\Bigg]\nonumber\\
&\kappa_{12}=-\frac{\Omega\sqrt{\Omega^2+\Omega^2_1}}{2\sqrt{2}\Omega'^2}\Bigg[\frac{\Omega'+i\Omega_1}{\Omega'-i\Omega_1}\Gamma_2-\frac{i2\Omega_1}{\Omega'-i\Omega_1}\Gamma_3-\Gamma_4\Bigg]\\
&\kappa_{13}=-\frac{\Omega^2+\Omega_1^2}{4\Omega'^2}\Big[-\Big(\frac{\Omega'+i\Omega_1}{\Omega'-i\Omega_1}\Big)^2\Gamma_2+\frac{4\Omega^2}{(\Omega'-i\Omega_1)^2}\Gamma_3-\Gamma_4\Big]\nonumber\\
&\kappa_{23}=-\frac{\Omega\sqrt{\Omega^2+\Omega^2_1}}{2\sqrt{2}\Omega'^2}\Bigg[\frac{\Omega'+i\Omega_1}{\Omega'-i\Omega_1}\Gamma_2-\frac{i2\Omega_1}{\Omega'-i\Omega_1}\Gamma_3-\Gamma_4\Bigg]\nonumber\\
&\Omega_{p_1}=\frac{\sqrt{\Omega_1^2+\Omega^2}}{\sqrt{2}\Omega'}\frac{\Omega'-i\Omega_1}{\Omega'+i\Omega_1}\Omega_{12};~\Omega_{p_2}=-\frac{\Omega}{\Omega'}\Omega_{12};\nonumber\\
&\Omega_{p_3}=\frac{\sqrt{\Omega_1^2+\Omega^2}}{\sqrt{2}\Omega'}\frac{\Omega'+i\Omega_1}{\Omega'-i\Omega_1}\Omega_{12};~\textrm{where, } \Omega'=\sqrt{2\Omega^2+\Omega_1^2}\nonumber
\end{align}
\normalsize{}
\begin{figure}
   \begin{center}
      \includegraphics[width =0.8\linewidth]{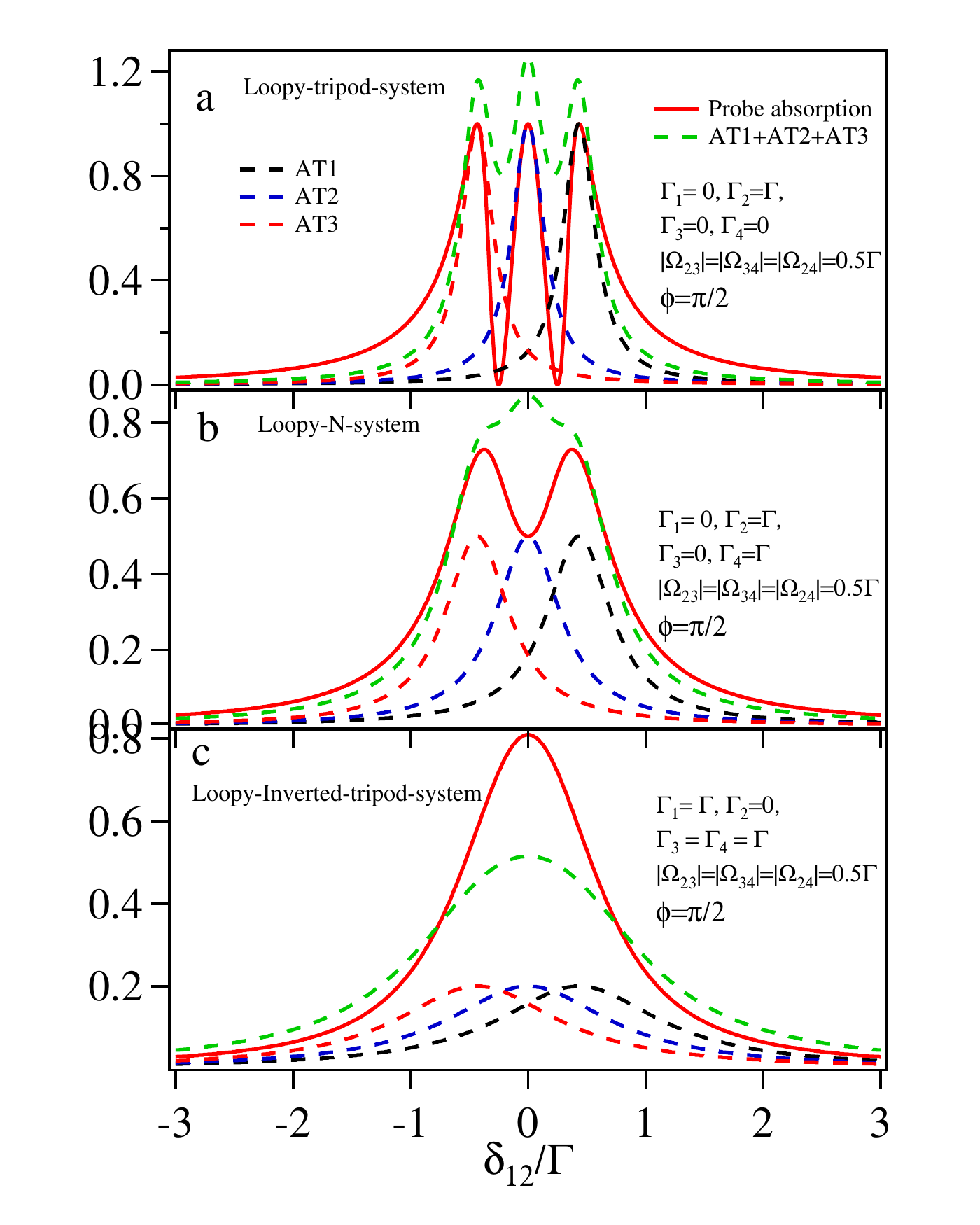}
 \caption{AT peaks and the effect of interference for probe absorption vs  its detuning ($\delta_{12}/\Gamma$) with $\delta_{23}=\delta_{34}=\delta_{24}=0$, $\Omega_{23}=\Omega_{34}=\Omega_{24}$ i.e. $\Omega_{p_1}=\frac{1}{\sqrt{3}}\frac{\sqrt{3}-i}{\sqrt{3}+i}\Omega_{12}$, $\Omega_{p_2}=-\frac{\Omega_{12}}{\sqrt{3}}$, $\Omega_{p_3}=\frac{1}{\sqrt{3}}\frac{\sqrt{3}+i}{\sqrt{3}-i}\Omega_{12}$ and $\phi=\pi/2$. a) $\kappa_{12}=\kappa_{23}=-\frac{\sqrt{3}+i}{\sqrt{3}-i}\frac{\Gamma}{6}$, $\kappa_{13}=+\left(\frac{\sqrt{3}+i}{\sqrt{3}-i}\right)^2\frac{\Gamma}{6}$, $\Gamma_{d_1}=\Gamma_{d_2}=\Gamma_{d_3}=\frac{\Gamma}{6}$ b)  $\kappa_{12}=\kappa_{23}=-\frac{i}{\sqrt{3}-i}\frac{\Gamma}{3}$, $\kappa_{23}=+\left[\left(\frac{\sqrt{3}+i}{\sqrt{3}-i}\right)^2+1\right]\frac{\Gamma}{6}$, $\Gamma_{d_1}=\Gamma_{d_2}=\Gamma_{d_3}=\frac{\Gamma}{3}$ c) $\kappa_{12}=\kappa_{23}=+\frac{\sqrt{3}+i}{\sqrt{3}-i}\frac{\Gamma}{6}$, $\kappa_{13}=-\left[\left(\frac{2}{\sqrt{3}-i}\right)^2-1\right]\frac{\Gamma}{6}$, $\Gamma_{d_1}=\Gamma_{d_2}=\Gamma_{d_3}=\frac{\Gamma}{3}$.}
      \label{Fig12}
   \end{center}
\end{figure}
We similarly consider the various combinations of $\Gamma_2,~\Gamma_3$ and $\Gamma_4$ to see the nature of interference between the AT peaks and take $\Omega_{23}=\Omega_{34}=\Omega_{24}=0.5\Gamma$ for all the cases of the loopy systems for $\phi=\pi/2$. 
\begin{enumerate}[(i)]
\item For $\Gamma_1=\Gamma_3=\Gamma_4=0$, $\Gamma_2=\Gamma$ (Loopy-tripod system): 

For this case, all the individual terms of ``Interference1" and ``Interference2" are negative and hence there is prominent destructive interference between the AT peaks (see Fig. \ref{Fig12}a), since the probe absorption goes to zero at the overlapping region of the AT peaks. 

\item For $\Gamma_2=\Gamma_4=\Gamma$, $\Gamma_1=\Gamma_3=0$ (loopy N-system):

For this case, the magnitude of the interference terms are less compared to the loopy-tripod system as shown in Fig. \ref{Fig12}b. 
 
\item For $\Gamma_1=\Gamma_3=\Gamma_4=\Gamma$, $\Gamma_2=0$ (loopy inverted-tripod system):

All the individual terms of the ``Interference1" and ``Interference2" are positive and hence there is constructive interference between the AT peaks as shown in Fig. \ref{Fig12}c.
\end{enumerate}

\section{Conclusion}

In conclusion, we present the theoretical frame work to identify the nature and the role of interference between the Autler-Townes peaks (dressed states) in multi-level system. In three-level system the two AT peaks  interferes pair-wise and the nature of interference is very simple which can be constructive, destructive or no interference depending upon the decay rate of the states coupled by the strong control lasers. In four-level system the nature of interference is more complicated but again all the three AT peaks interferes pair-wise. There are two terms for the interference, one that is similar to the three-level system and another that is a little bit more complicated. We have also done similar calculation for five level system which is involves further complicated nature of the interference. For any system, if the decay rate of the levels coupled by the control lasers are equal then there is no interference between any of the AT peaks.   
\section{Acknowledgment}
E.O.N. would like to acknowledge Indian Council for Cultural Relations (ICCR) for the PhD scholarship. K.P. would like to acknowledge the funding from SERB of grant No. ECR/2017/000781. We would like to thank David Wilkowski for his intriguing comments. 

\section*{Author contribution statement}

KP conceived the idea including the formulation of the theoretical concept and drafting of the manuscript. EON has cross checked all the calculations including preparing the manuscript. DS and KI have partially contributed to this manuscript. All the authors have read and approved the final version of the manuscript. 

\bibliographystyle{epj}
\bibliography{eitrefsall}

\begin{thebibliography}{39}

\bibitem{IMA89}
A.~Imamog\ifmmode~\check{}\else \v{}\fi{}lu, Phys. Rev. A \textbf{40}, 2835
  (1989)

\bibitem{LIX95}
Y.q. Li, M.~Xiao, Phys. Rev. A \textbf{51}, 4959 (1995)

\bibitem{FIM05}
M.~Fleischhauer, A.~Imamoglu, J.P. Marangos, Rev. Mod. Phys. \textbf{77}, 633
  (2005)

\bibitem{SHF96}
S.~Shepherd, D.J. Fulton, M.H. Dunn, Phys. Rev. A \textbf{54}, 5394 (1996)

\bibitem{AGA97}
G.S. Agarwal, Phys. Rev. A \textbf{55}, 2467 (1997)

\bibitem{ABT10}
T.Y. Abi-Salloum, Phys. Rev. A \textbf{81}, 053836 (2010)

\bibitem{TAH14}
C.~Tan, G.~Huang, J. Opt. Soc. Am. B \textbf{31}, 704 (2014)

\bibitem{KBN16}
S.~Khan, V.~Bharti, V.~Natarajan, Physics Letters A \textbf{380}, 4100  (2016)

\bibitem{ADS11}
P.M. Anisimov, J.P. Dowling, B.C. Sanders, Phys. Rev. Lett. \textbf{107},
  163604 (2011)

\bibitem{GVS13}
L.~Giner, L.~Veissier, B.~Sparkes, A.S. Sheremet, A.~Nicolas, O.S. Mishina,
  M.~Scherman, S.~Burks, I.~Shomroni, D.V. Kupriyanov et~al., Phys. Rev. A
  \textbf{87}, 013823 (2013)

\bibitem{HJX18}
L.~Hao, Y.~Jiao, Y.~Xue, X.~Han, S.~Bai, J.~Zhao, G.~Raithel, New Journal of
  Physics \textbf{20}, 073024 (2018)

\bibitem{LWZ20}
J.~Liu, J.~Wu, Y.~Zhang, Y.~He, J.~Zhang, J. Opt. Soc. Am. B \textbf{37}, 49
  (2020)

\bibitem{POC14}
B.~Peng, Å.K. Ozdemir, W.~Chen, F.~Nori, L.~Yang, \textbf{5} (2014)

\bibitem{ETA18}
E.~Saglamyurek, T.~Hrushevskyi, A.~Rastogi, K.~Heshami, L.J. LeBlanc, Nature
  Photonics \textbf{12}, 774  (2018)

\bibitem{RSH19}
A.~Rastogi, E.~Saglamyurek, T.~Hrushevskyi, S.~Hubele, L.J. LeBlanc, arXiv
  preprint arXiv:1902.02815  (2019)

\bibitem{SCJ10}
U.~Sinha, C.~Couteau, T.~Jennewein, R.~Laflamme, G.~Weihs, Science
  \textbf{329}, 418 (2010),
  \texttt{http://science.sciencemag.org/content/329/5990/418.full.pdf}

\bibitem{ASQ15}
M.~Asad~Siddiqui, T.~Qureshi, Progress of Theoretical and Experimental Physics
  \textbf{2015} (2015),
  \texttt{http://oup.prod.sis.lan/ptep/article-pdf/2015/8/083A02/7698190/ptv112.pdf}

\bibitem{SAS15}
A.~Sinha, A.~H.~Vijay, U.~Sinha, Scientific Reports \textbf{5}, 10304 (2015)

\bibitem{LZC20}
K.S. Lee, Z.~Zhuo, C.~Couteau, D.~Wilkowski, T.~Paterek, Phys. Rev. A
  \textbf{101}, 052111 (2020)

\bibitem{JOX03}
A.~Joshi, M.~Xiao, Physics Letters A \textbf{317}, 370  (2003)

\bibitem{BMW08}
M.G. Bason, A.K. Mohapatra, K.J. Weatherill, C.S. Adams, Journal of Physics B:
  Atomic, Molecular and Optical Physics \textbf{42}, 075503 (5pp) (2009)

\bibitem{CXH09}
Y.~Chen, X.G. Wei, B.S. Ham, Journal of Physics B: Atomic, Molecular and
  Optical Physics \textbf{42}, 065506 (2009)

\bibitem{SYK11}
J.~Sheng, X.~Yang, U.~Khadka, M.~Xiao, Opt. Express \textbf{19}, 17059 (2011)

\bibitem{HCN05}
T.~Hong, C.~Cramer, W.~Nagourney, E.N. Fortson, Phys. Rev. Lett. \textbf{94},
  050801 (2005)

\bibitem{JSA12}
J.A. Sedlacek, A.~Schwettmann, H.~Kubler, R.~Low, T.~Pfau, J.P. Shaffer, Nature
  physics \textbf{8}, 819 (2012)

\bibitem{BJX11}
B.~Zhang, J.H. Wu, X.Z. Yan, L.~Wang, X.J. Zhang, J.Y. Gao, Opt. Express
  \textbf{19}, 12000 (2011)

\bibitem{GIM18}
A.~Ghosh, K.~Islam, S.~Mondal, D.~Bhattacharyya, N.~Pal, A.~Bandyopadhyay,
  Journal of Physics B: Atomic, Molecular and Optical Physics \textbf{51},
  145501 (2018)

\bibitem{SLB13}
S.~Kumar, T.~Laupr\^etre, F.~Bretenaker, F.~Goldfarb, R.~Ghosh, Phys. Rev. A
  \textbf{88}, 023852 (2013)

\bibitem{LPR18}
F.~Leroux, K.~Pandey, R.~Rehbi, F.~Chevy, C.~Miniatura, B.~Gremaud,
  D.~Wilkowski, Nature Communications \textbf{9}, 3580 (2018)

\bibitem{HMW14}
Y.X. Hu, C.~Miniatura, D.~Wilkowski, B.~Gr\'emaud, Phys. Rev. A \textbf{90},
  023601 (2014)

\bibitem{PKP16}
K.~Pandey, C.C. Kwong, M.S. Pramod, D.~Wilkowski, Phys. Rev. A \textbf{93},
  053428 (2016)

\bibitem{LHF05}
L.~Li, H.~Guo, F.~Xiao, X.~Peng, X.~Chen, J. Opt. Soc. Am. B \textbf{22}, 1309
  (2005)

\bibitem{MOS83}
J.R. Morris, B.W. Shore, Phys. Rev. A \textbf{27}, 906 (1983)

\bibitem{PAN13}
K.~Pandey, Phys. Rev. A \textbf{87}, 043838 (2013)

\bibitem{CQB05}
I.~Courtillot, A.~Quessada-Vial, A.~Brusch, D.~Kolker, G.D. Rovera, P.~Lemonde,
  The European Physical Journal D - Atomic, Molecular, Optical and Plasma
  Physics \textbf{33}, 161 (2005)

\bibitem{MDP17}
N.S. Mallick, T.N. Dey, K.~Pandey, arXiv:1703.10492  (2017)

\bibitem{SOP18}
D.~Shylla, E.N. Ogaro, K.~Pandey, Scientific Reports \textbf{8}, 8692 (2018)

\bibitem{SHP18}
D.~Shylla, K.~Pandey, arXiv:1802.09935  (2018)

\bibitem{TDL18}
M.T. Simons, M.D. Kautz, C.L. Holloway, D.A. Anderson, G.~Raithel, D.~Stack,
  M.C. St.~John, W.~Su, Journal of Applied Physics \textbf{123}, 203105 (2018),
  \texttt{https://doi.org/10.1063/1.5020173}

\end{thebibliography}

\end{document}